\documentclass[twoside,twocolumn,9pt]{article}
\usepackage{extsizes}
\usepackage[super,sort&compress,comma]{natbib} 
\usepackage[version=3]{mhchem}
\usepackage[left=1.5cm, right=1.5cm, top=1.785cm, bottom=2.0cm]{geometry}
\usepackage{balance}
\usepackage{times,mathptmx}
\usepackage{sectsty}
\usepackage{graphicx} 
\usepackage{lastpage}
\usepackage[format=plain,justification=justified,singlelinecheck=false,font={stretch=1.125,small,sf},labelfont=bf,labelsep=space]{caption}
\usepackage{float}
\usepackage{fancyhdr}
\usepackage{fnpos}
\usepackage[english]{babel}
\addto{\captionsenglish}{%
  
}
\usepackage{array}
\usepackage{droidsans}
\usepackage{charter}
\usepackage[T1]{fontenc}
\usepackage[usenames,dvipsnames]{xcolor}
\usepackage{setspace}
\usepackage[compact]{titlesec}
\usepackage{hyperref}
\usepackage{epstopdf}%This line makes .eps figures into .pdf - please comment out if not required.

\definecolor{cream}{RGB}{222,217,201}

\begin{document}

\pagestyle{fancy}
\thispagestyle{plain}
\fancypagestyle{plain}{

%%%HEADER%%%
\renewcommand{\headrulewidth}{0pt}
}
%%%END OF HEADER%%%

%%%PAGE SETUP - Please do not change any commands within this section%%%
\makeFNbottom
\makeatletter
\renewcommand\LARGE{\@setfontsize\LARGE{15pt}{17}}
\renewcommand\Large{\@setfontsize\Large{12pt}{14}}
\renewcommand\large{\@setfontsize\large{10pt}{12}}
\renewcommand\footnotesize{\@setfontsize\footnotesize{7pt}{10}}
\makeatother

\renewcommand{\thefootnote}{\fnsymbol{footnote}}
\renewcommand\footnoterule{\vspace*{1pt}% 
\color{cream}\hrule width 3.5in height 0.4pt \color{black}\vspace*{5pt}} 
\setcounter{secnumdepth}{5}

\makeatletter 
\renewcommand\@biblabel[1]{#1}            
\renewcommand\@makefntext[1]% 
{\noindent\makebox[0pt][r]{\@thefnmark\,}#1}
\makeatother 
\renewcommand{\figurename}{\small{Fig.}~}
\sectionfont{\sffamily\Large}
\subsectionfont{\normalsize}
\subsubsectionfont{\bf}
\setstretch{1.125} %In particular, please do not alter this line.
\setlength{\skip\footins}{0.8cm}
\setlength{\footnotesep}{0.25cm}
\setlength{\jot}{10pt}
\titlespacing*{\section}{0pt}{4pt}{4pt}
\titlespacing*{\subsection}{0pt}{15pt}{1pt}
%%%END OF PAGE SETUP%%%

%%%FOOTER%%%
\fancyfoot{}
\fancyfoot[LO,RE]{\vspace{-7.1pt}\includegraphics[height=9pt]{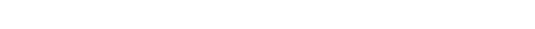}}
\fancyfoot[CO]{\vspace{-7.1pt}\hspace{13.2cm}\includegraphics{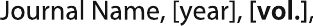}}
\fancyfoot[CE]{\vspace{-7.2pt}\hspace{-14.2cm}\includegraphics{head_foot/RF}}
\fancyfoot[RO]{\footnotesize{\sffamily{1--\pageref{LastPage} ~\textbar  \hspace{2pt}\thepage}}}
\fancyfoot[LE]{\footnotesize{\sffamily{\thepage~\textbar\hspace{3.45cm} 1--\pageref{LastPage}}}}
\fancyhead{}
\renewcommand{\headrulewidth}{0pt} 
\renewcommand{\footrulewidth}{0pt}
\setlength{\arrayrulewidth}{1pt}
\setlength{\columnsep}{6.5mm}
\setlength\bibsep{1pt}
%%%END OF FOOTER%%%

%%%FIGURE SETUP - please do not change any commands within this section%%%
\makeatletter 
\newlength{\figrulesep} 
\setlength{\figrulesep}{0.5\textfloatsep} 

\newcommand{\topfigrule}{\vspace*{-1pt}% 
\noindent{\color{cream}\rule[-\figrulesep]{\columnwidth}{1.5pt}} }

\newcommand{\botfigrule}{\vspace*{-2pt}% 
\noindent{\color{cream}\rule[\figrulesep]{\columnwidth}{1.5pt}} }

\newcommand{\dblfigrule}{\vspace*{-1pt}% 
\noindent{\color{cream}\rule[-\figrulesep]{\textwidth}{1.5pt}} }

\makeatother
%%%END OF FIGURE SETUP%%%

%%%TITLE, AUTHORS AND ABSTRACT%%%
\twocolumn[
  \begin{@twocolumnfalse}
{\includegraphics[height=30pt]{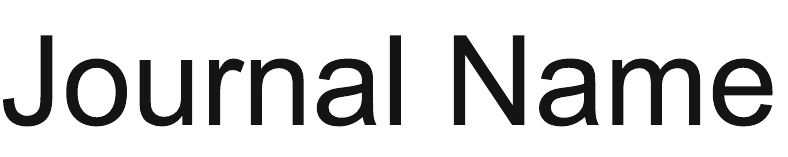}\hfill%
 \raisebox{0pt}[0pt][0pt]{\includegraphics[height=55pt]{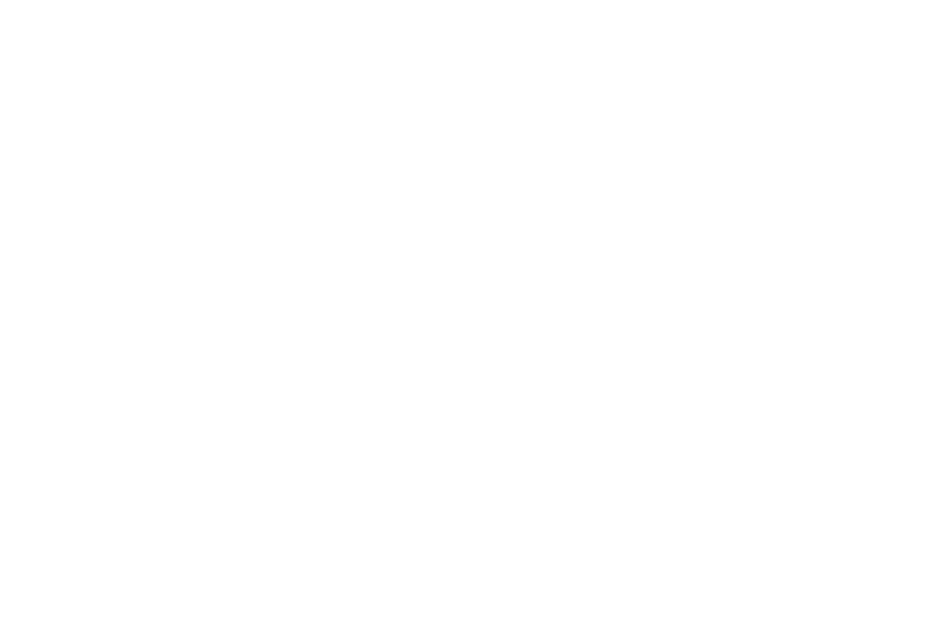}}%
 \\[1ex]%
 \includegraphics[width=18.5cm]{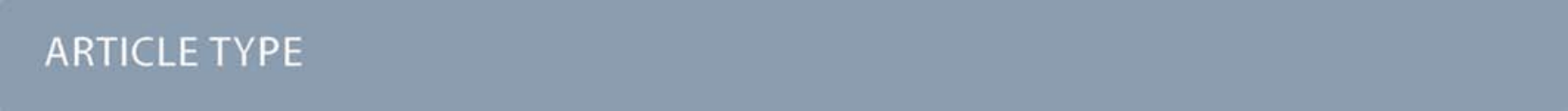}}\par
\vspace{1cm}
\sffamily
\begin{tabular}{m{4.5cm} p{13.5cm} }

\includegraphics{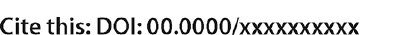} & \noindent\LARGE{\textbf{A high throughput search of efficient thermoelectric half-Heusler compounds$^\dag$}} \\%Article title goes here instead of the text "This is the title"
\vspace{0.3cm} & \vspace{0.3cm} \\

 & \noindent\large{Parul R. Raghuvanshi\textit{$^{a}$}, Suman Mondal\textit{$^{a}$}, and Amrita Bhattacharya$^{\ast}$\textit{$^{a}$}} \\%Author names go here instead of "Full name", etc.

\includegraphics{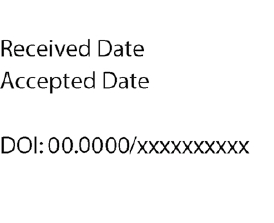} & \noindent\normalsize{Half-Heusler compounds have emerged as promising thermoelectric materials that offer huge compositional space to tune their thermoelectric performance. A class of stable half Heusler compounds formed from elements of three specific groups in the periodic table viz. X$_{p}$X$'_{1-p}$Y$_{q}$Y$'_{1-q}$Z$_{r}$Z$'_{1-r}$ (with X, X$'$= Ti, Zr, Hf, Y, Y$'$ = Ni, Pd, Pt and Z, Z$'$ = Ge, Sn, Pb and p, q, r = 0, 0.25, 0.75 and 1) via various stoichimetric isoelectronic elemental substitution at the X, Y and Z sites respectively is investigated. Intelligent filters are employed at each step of our high throughput density functional theory calculations to filter compounds with improved figure of merit. While confirming several known results, the calculations also reveal unknown pathways to improve the thermoelectric performance of the compound class. The 50\% X as well as Z site substitution of the parent Heusler individually are found to marginally enhance the power factor for both the $p$- and $n$-type doping, while leading to considerable enhancement in the figure of merit (by $\sim$24 \%) specifically due to lowering of the lattice thermal conductivity because of increase in lattice disorder in approximately the same cell volume. Furthermore, the present study confirms the experimental scenario that Y site substitution does not lead to enhancement of the powerfactor because of the breaking of band degeneracies at the high symmetry points. This work will serve as a consolidated cost effective guideline for experimentalist working with this compound class on enhancing the powerfactor and figure of merit of the compositions.} \\%The abstrast goes here instead of the text "The abstract should be..."

\end{tabular}

 \end{@twocolumnfalse} \vspace{0.6cm}

  ]
%%%END OF TITLE, AUTHORS AND ABSTRACT%%%

%%%FONT SETUP - please do not change any commands within this section
\renewcommand*\rmdefault{bch}\normalfont\upshape
\rmfamily
\section*{}
\vspace{-1cm}

%%%FOOTNOTES%%%

\footnotetext{$^{\ast}$\textit{$^{a}$~Department of metallurgical engineering and materials science, Indian Institute of Technology, Bombay, Powai-400076, Mumbai, Maharashtra, India. E-mail: b\_amrita@iitb.ac.in}}

%Please use \dag to cite the ESI in the main text of the article.
%If you article does not have ESI please remove the the \dag symbol from the title and the footnotetext below.
\footnotetext{\dag~Electronic Supplementary Information (ESI) available: [details of any supplementary information available should be included here]. See DOI: 00.0000/00000000.}
%additional addresses can be cited as above using the lower-case letters, c, d, e... If all authors are from the same address, no letter is required

%%%END OF FOOTNOTES%%%

%%%MAIN TEXT%%%%
\section{Introduction}

One of the promising alternative to meet the ever increasing energy demands of the modern world is the large scale deployment of energy efficient thermoelectric devices, which can convert a temperature gradient into an electric voltage~\cite{DiSalvo1999,Tritt1999,Bell2008}. However, the main challenge in this development lies with the low efficiency of the thermoelectric materials, which inturn depends on the dimensionless figure of merit $zT$ of the material~\cite{Snyder2008}. $zT$ of a material at a given operating temperature depends on the instrinsic parameters i.e. the Seebeck $S$, the electrical conductivity $\sigma$, and the total thermal conductivity $\kappa$ via the expression $zT= \frac{S^{2}\sigma T}{\kappa}$~\cite{Liu2015,QZhang2016}. The numerator is collectively known as power factor ($S^{2}\sigma$), which is contributed by the electronic transport coefficients, whereas, the denominator $\kappa$ is the sum of the heat generated by transportation of: (a) charge carriers (i.e. electron and hole) comprising the electronic thermal conductivity ($\kappa_{el}$) and (b) phonons travelling between the lattices comprising the lattice thermal conductivity ($\kappa_{l}$). Naturally, the main strategies that are generally employed to optimize the $zT$ either involve maximizing the power factor in the numerator or mininmizing the $\kappa_{l}$ in the denominator~(which may be indepedently tuned with respect to the electronic transport coefficients in many materials) or both~\cite{Kimura2009,CFu2015,Holuj2015,TJZhu2017,Amrita2020}. The electronic transport coefficients can be tuned by band engineering viz. by tuning the degeneracy of the valence/ conduction band valleys, the effective mass of the carriers, the scattering lifetimes and carrier concentration~\cite{Goldsmid1964,Caillat2000,XYan2012,Pei2012,HXie2013,Nagendra2020}. The $\kappa_{l}$ on the other hand can be reduced by shortening the phonon mean-free-path, which can be achieved by introducing mass disorder in the lattice~\cite{Toberer2012, loffe1960}. Thus, analysis of the electronic structure via substitution or doping has often been found to provide critical insights and in recent years, band engineering via first principles calculations has proven to be a cost effective and efficient way for tuning these parameters prior to their synthesis in the laboratory~\cite{Sandip2016}. High throughput calculations of large number of compounds have been recently carried out to screen the ones with desirable properties for their thermoelectric applications~\cite{Carrete2014,JCarrete2014,Guo2019}.

Half Heuslers (HH) are intermetallic ternary compounds (with formula unit XYZ), where X and Y are transition or rare earth metal, while Z is a p-block element. They comprise one of the most promising, earth abundant and environment friendly class of thermoelectric material offering a huge chemical space to tune the transport coefficients~\cite{JYang2008,WXie2012,SChen2013,TZhu2015,Zeier2016}. X and Y atoms are distinct in cationic character, whereas Z is an anionic counterpart. They crystallize in a noncentrosymmetric lattice (space group F$\bar{4}$3m), which has four symmetry unique positions for accommodating the atoms; X (0, 0, 0), Z (1/2, 1/2, 1/2), Y (1/4, 1/4, 1/4), and a vacant fourth position (3/4, 3/4, 3/4)~\cite{}. If the fourth vacant position is also occupied by Y, the full Heusler (FH) alloy with the general formula XY$_2$Z (space group Fm$\bar{3}$m) is formed \cite{}. HH may be semiconducting or metallic depending upon the total valency of the elements in the compound. HH compounds, which have 8 (I-I-VI, I-II-V, I-III-IV, II-II-IV, and II-III-III) and 18 (XI-I-VI, XI-II-V, XI-III-IV, I-XII-V, II-XII-IV, III-XII-III, X-II-VI, X-III-V, and X-IV-IV), generally show semiconducting nature with narrow band gap~\cite{Aliev1989,Aliev1990,Pierre1997,Larson1999,Ciftci2016,HZhu2019}.     

Naturally, many different chemical substitutions as well as dopings have been tried in the past leading to the increase in the $zT$~\cite{Ogut1995,Joshi2011,HXie2014,GRogl2017} upto 1.5, which is so far the highest reported value in this class~\cite{CFu2015, GRogl2017, Poon2018}. Hohl and coworkers~\cite{Hohl1999} explained the effect of various sites substitution and doping on the $zT$ of these half Heulser. They showed that X site substitution improves the power factor in Ti$_{0.5}$Hf$_{0.5}$NiSn by 1.18 times as compared to pristine HfNiSn (which has a power factor of 1.7 $\mu$WK$^{-2}$cm$^{-1}$). They also reported that doping by Nb at the X site, further improves the power factor to 13 times by containing 1\% Nb composition. Later, Sakurada and Shutoh~\cite{Sakurada2005} also showed the improvement of $zT$ due to the combined effect of high power factor and low thermal conductivity by the X and Z site substitutions and obtained $zT$ of 1.3 at 700 K for (Zr$_{0.5}$Hf$_{0.5}$)$_{0.5}$Ti$_{0.5}$NiSn and 1.5 for further Sn doped (Zr$_{0.5}$Hf$_{0.5}$)$_{0.5}$Ti$_{0.5}$NiSn$_{0.998}$Sb$_{0.002}$ at the same temperature. Zhang \textit{et. al.}~\cite{Zhang2017} used first principle calculations to explain the reasons behind the increase in the Seebeck coefficient for X site substitution and power factor improvement due to Z site co-doping. Zou \textit{et al.} and other independent groups investigated the maximum power factor and corresponding optimal carrier concentration due to 50\% Hf substitution at Zr site~\cite{Yang2002, DZou2013} using first principle methods and suggested guidelines for tuning the doping levels and composition to enhance the thermoelectric performance~\cite{Chaput2006,Downie2014,Downie2015,Rogl2019}. Each of these studies provides some critical explanation towards the step by step realization of high $zT$ HH thermoelectrics and reveals the huge potential of this compound class. While the vast majority of earlier reported studies have focussed on analysing the effect of tuning the power factor, the main challenge in implementation of half Heusler compounds as thermoelectric arises from their large value of $\kappa_{l}$, inspite of their tunable narrow band gap and high power factor. Hence, it is supremely important to explore the scope of reducing the $\kappa_{l}$ via introducing mass disorder, without compromising on the vital parameters in the electronic structure leading to its high power factor~\cite{YLiu2015,Chauhan2019,Anand2019}. If such compounds can be identified they may act as the perfect basis for chemical doping for tuning the carrier concentration and hence, improving the $zT$.     

In this article, the effect of compositional variation via isoelectronic substitution is studied on the stability, electronic and transport properties of semiconducting X$_{p}$X$'_{1-p}$Y$_{q}$Y$'_{1-q}$Z$_{r}$Z$'_{1-r}$ (X, X$'$= Ti, Zr, Hf; Y, Y$'$ = Ni, Pd, Pt and Z, Z$'$ = Ge, Sn, Pb) HH compositions is explored, where $p$, $q$, and $r$ are the different concentrations of elemental isoelectronic substitution at the X, Y and Z sites respectively. The purpose of this work is to identify compound(s) that satisfy the simultaneous requisite criteria of high power factor and low thermal conductivity, so that it may act as a semiconducting basis for further chemical doping.   

\begin{figure}
\centering
\includegraphics[scale=0.325]{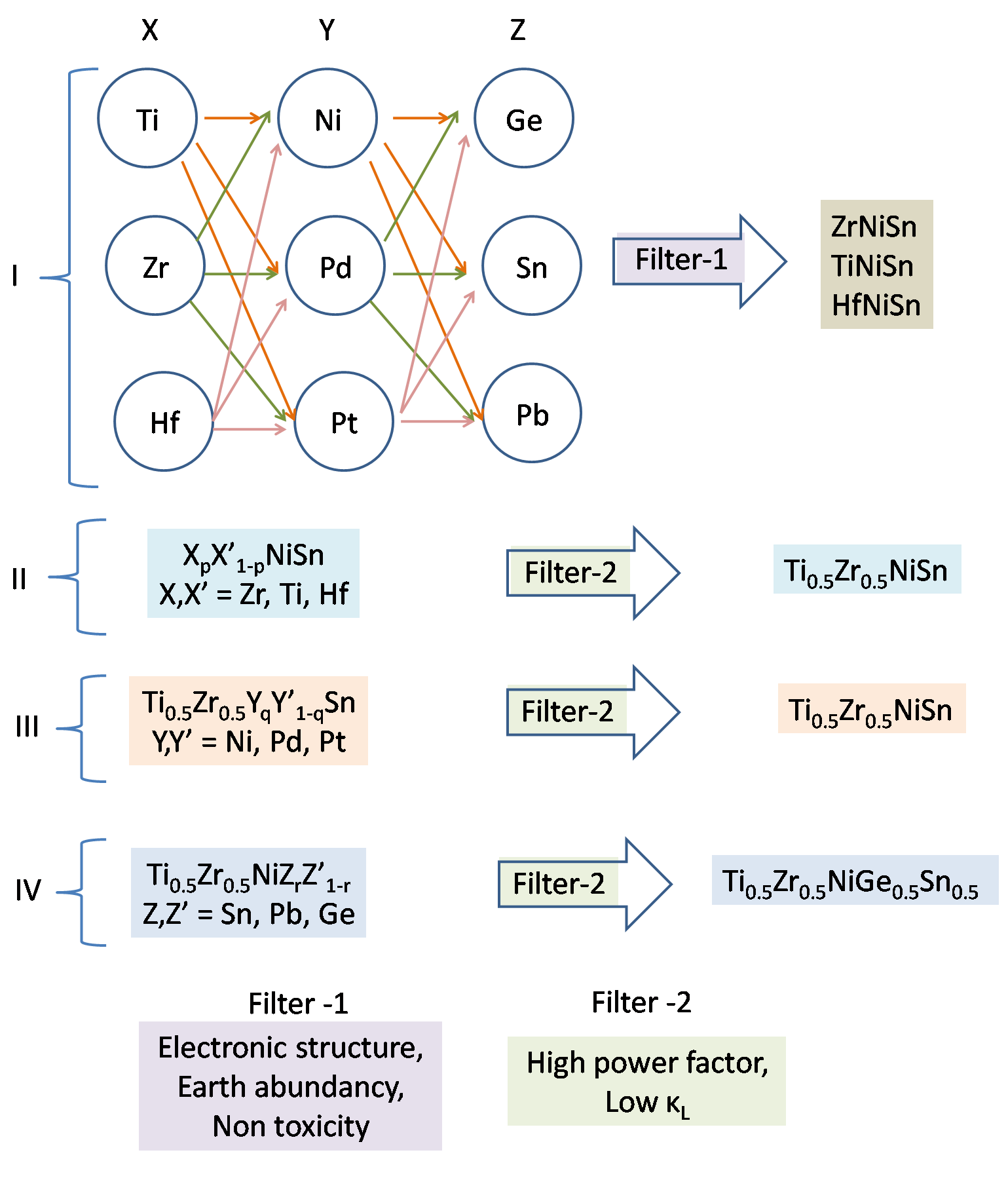}
\caption{Work flow for achieving the X$_{p}$X$'_{1-p}$Y$_{q}$Y$'_{1-q}$Z$_{r}$Z$'_{1-r}$ composition with improved thermoelectric efficiency. Different filters are applied at different steps to screen the compounds~(see text for details).} 
\label{fig:Highthruput}
\end{figure}

\section{Computational Details}
High throughput density functional theory~(DFT)~\cite{Hohenberg1964, Kohn1965} calculations are performed using the Vienna Ab-initio Simulation Program (VASP)~\cite{Kresse1994, Hafner1997}, which is a plane wave-based electronic structure code with Projector Augmented Wave (PAW) and Perdew-Burke-Ernzerhof (PBE)~\cite{PBE1996} formulation of the exchange-correlation energy functional under a Generalized Gradient Approximation (GGA). Plane-Wave cutoff energy of 500 eV and an energy convergence criterion of {\bf 10$^{-3}$ eV} are used. For static calculations, a coverged Monkhorst-Pack k-mesh grid is applied for the unit cell. For each structure, ionic as well as geometric relaxations are performed. 

In order to examine the stability of the compositions, the formation energy of the compounds is calculated as;

\begin{equation}
\label{Eform}
E_\mathrm{f} = E(X_{m}Y_{n}Z_{t})-m \cdot \frac{E(X_{a})}{a}-n \cdot \frac{E(Y_{b})}{b}- t \cdot \frac{E(Z_{c})}{c}
\end{equation}
where, $E(X_{m}Y_{n}Z_{t})$ is the total energy of the compound and $E(X_{a})$, $E(Y_{b})$, and $E(Z_{c})$ are the total energy of the bulk phases of X, Y, and Z with $a$, $b$ and $c$ numbers of atoms respectively. 

The electronic transport coefficients i.e. the thermopower ($S$) and electronic conductivity ($\sigma$) are calculated from the semi-classical Boltzmann theory~\cite{Ziman1960} as implemented in the BoltzTraP package~\cite{MADSEN200667} where, scattering is treated perturbatively (but not self-consistently) on top of the rigid band and constant relaxation time approximation. Accordingly, even the thermal renormalization of the electronic structure, i.e., the temperature dependent change of the electronic band structure due to the electron-phonon coupling, is typically neglected in these approaches~\cite{Sam2018}. However, this pretty brute simplification implies huge savings in the computational effort and thus enables large-scale, high-throughput calculations. Fully optimized crystal structures are considered for the transport calculations, whereby a dense Monkhorst-Pack K-mesh is used for obtaining the energy eigen values.

The phonon band structure is calculated using the finite displacement method as implemented in the phonopy~\cite{Togo2015} code. Converged supercell of 2$\times$2$\times$2 is used to calculate the phonon band structure. The amplitude of the displacements is fixed to 0.01 \AA, and the forces are converged to the accuracy of 10$^{-8}$eV/\AA. The phonon group velocity $v$ ($v= \mathrm{d\omega}/\mathrm{dK}$), mode resolved phonon group velocity ${v_i}$~($v_i= \mathrm{d\omega}_i/\mathrm{dK}$), Gr$\ddot{\mathrm{u}}$neisen parameter $\gamma$~($\gamma = -\frac{\mathrm{V_0}}{\omega}\frac{\mathrm{d\omega}}{\mathrm{dV}}$), and mode resolved Gr$\ddot{\mathrm{u}}$neisen parameter $\gamma_i$ ($\gamma_i= -\frac{\mathrm{V_0}}{\omega_i}\frac{\mathrm{d\omega_i}}{\mathrm{dV}}$) are extracted from the harmonic phonon band dispersion using python based extensions to phonopy-VASP. For calculating the $\gamma$ and $\gamma_i$, the lattice is subjected to strains of $\pm$ 2~\%. Finally, we use the Asen-Palmer modified version of the Debye Callaway theory~\cite{Callaway1959} parameterized for solids~\cite{Asen1997,YZhang2012,NSChauhan2018,Amrita2018,Amrita2019,Amrita2020} to calculate $\kappa_{l}$ of the compounds.\\

\section{Result and Discussions}

\subsection{Compositional space}
The compositional space attempted in this work is enormous i.e. X$_{p}$X$'_{1-p}$Y$_{q}$Y$'_{1-q}$Z$_{r}$Z$'_{1-r}$ (where X, X$'$= Ti, Zr, Hf; Y, Y$'$ = Ni, Pd, Pt and Z, Z$'$ = Ge, Sn, Pb and $p$, $q$, and $r$ = 0, 0.25, 0.5, 0.75, 1). In order to reduce the number of compositions judicially without elimininating the vital ones, an intelligent screening policy is adopted. As the first step, the electronic structure of all the 27 parent compounds (i.e. those with one element occupying one site to attain the formula XYZ with X = Ti, Zr, Hf; Y = Ni, Pd, Pt, and Z= Ge, Sn, Pb) is computationally evaluated. The computed band gap of the parent compounds show wide variation i.e. from 0.35 to 1.12 eV (see supplimentary information for details). The compounds are preliminarily screened on the basis of their electronic band gap and number of valley degeneracy at the conduction band minimum (CBm) and valence band maximum (VBM), non toxicity and earth abundancy of the elemental constituents (as discussed in details in section \ref{sec:single}). After careful evaluation of the band structure ZrNiSn, TiNiSn and HfNiSn are selected with earth abundant and non toxic Ni and Sn occupying the Y and Z sites. The different stoichiometries (with varying $p$) for the composition X$_{p}$X$'_{1-p}$NiSn, with X, X$'$= Ti, Zr, Hf, is considered as the second step~(as discussed in details in section \ref{sec:X}). Once the stoichiometry and the composition for the X site substitution are identified on the basis of their high power factor and low lattice thermal conductivity, they are kept fixed for the subsequent Y site substitution. Thus, for the Y site substitution, the stoichiometry and the composition at the Y site is varried, but the previously identified compositions at the X site and the elemental composition at Z site is kept unchanged~(as discussed in section \ref{sec:Y} and \ref{sec:Z}). Finally, the same procedure is carried out for the Z site substitution, retaining the composition and stoichiometry for the X and Y site. The entire worflow is depicted schematically in Fig.~\ref{fig:Highthruput}, while the results are discussed in the suceeding sections.      
     
\subsection{Structure and stoichiometry}
\begin{figure}
\centering
\includegraphics[scale=0.36]{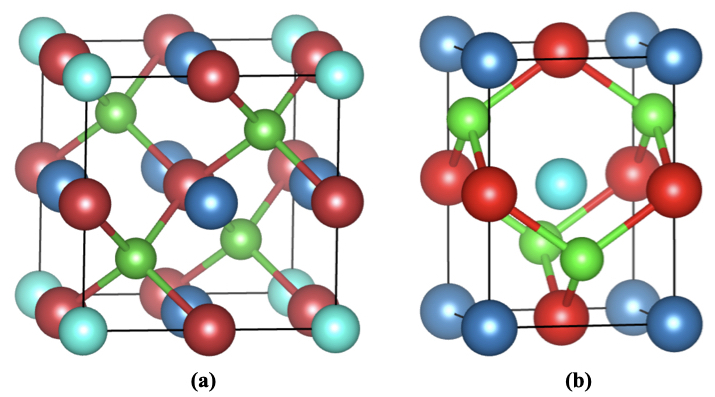}
\caption{Structure of the primitive cell of (a) 25\% (or 75\%) substituted X$_{0.25}$X$'_{0.75}$YZ (or X$_{0.75}$X$'_{0.25}$YZ) and (b) 50\% substituted X$_{0.5}$X$'_{0.5}$YZ half Heusler compound, where X, X$'$, Y, and Z atoms are shown in blue, cyan, green and red colors respectively.}
\label{Crystal} 
\end{figure}

For the sake of computational brevity, substitutions have been restricted to achieve three different stoichiometries apart from the parental compositions, viz. $p/ q/ r = 0.25, 0.5, 0.75$ to achieve the compositions X$_{p}$X$'_{1-p}$Y$_{q}$Y$'_{1-q}$Z$_{r}$Z$'_{1-r}$. The primitive cell (space group 216, $F\bar{4}3m$) is considered for calculating the electronic structure and transport coefficients of the parental HH compositions (with lattice parameter ranging from $\sim$ 4.0 - 4.5~\AA depending upon the variation in the compositional space). The conventional cubic lattice of XYZ (space group $F\bar{4}3m$) has 12 atoms (i.e. 4 X, 4 Y and 4 Z atoms) in the lattice (with the lattice parameter varying from $\sim$ 5.9 - 6.4 \AA). This particular lattice (i.e. 4~X, 4~Y, and 4~Z atoms per f.u) is used as the base structure to identify the ground state structure for the substitutions $p/ q/ r = 0.25, 0.75$ at different X, Y and Z site substituted cases, via an iterative scanning strategy. It is to be noted that such substitutions do not break the symmetry of the conventional cubic structure~(cf.~Fig.~\ref{Crystal}(a)). However, for $p/ q/ r = 0.5$, the primitive cell is found to have a tetragonal geometry with elongated \textit{c} axis and half the number of atoms (i.e. six) as in the conventional cell~(space group number 115, $P\bar{4}m2$ as shown in Fig.~\ref{Crystal}(b)). As for example, the elongated \textit{c} axis ($\sim$6.06 \AA) of 50\% substituted compound Ti$_{0.5}$Zr$_{0.5}$NiSn (tetragonal primitive cell) is found to lie between the lattice constant of ZrNiSn ($\sim$6.1 \AA) and TiNiSn ($\sim$5.9 \AA), which is in confirmation with Vegard's law\cite{Vegards1991,DZou2013}.

\subsection{Stability analysis}
\begin{figure}
\centering
\includegraphics[scale=0.36]{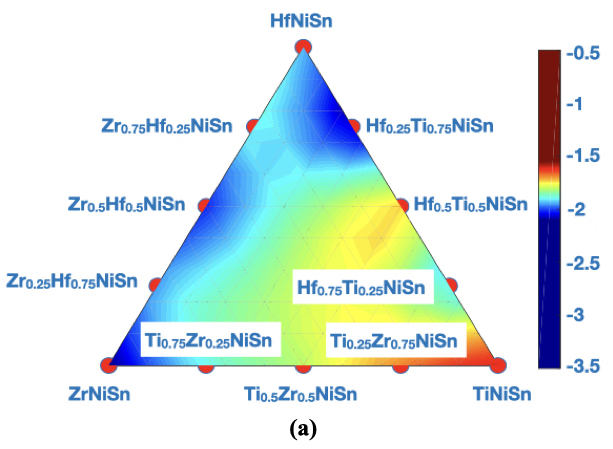}
\includegraphics[scale=0.34]{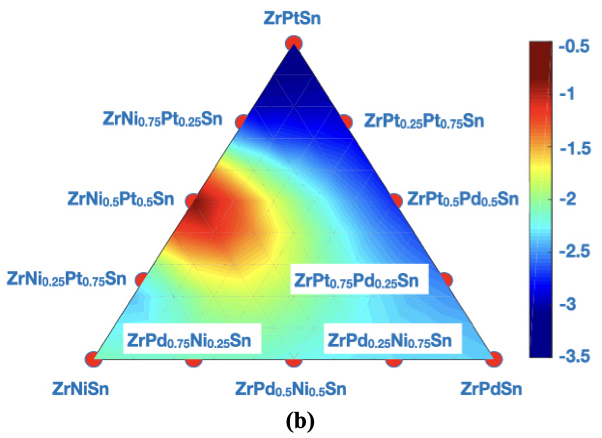}
\includegraphics[scale=0.33]{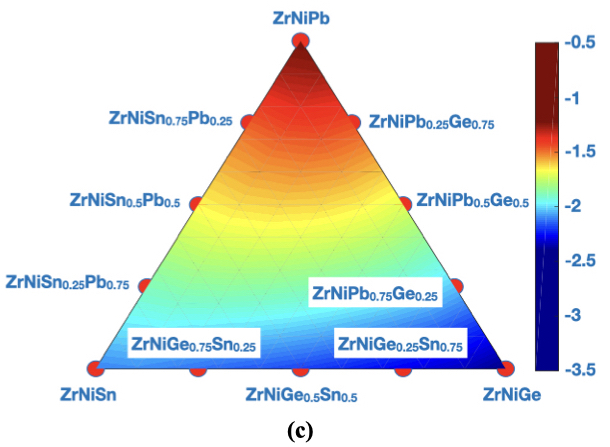}
\caption{Formation energy of X$_{p}$X$'_{1-p}$Y$_{q}$Y$'_{1-q}$Z$_{r}$Z$'_{1-r}$ for $p, q, r = 0, 0.25, 0.5, 0.75, 1$, calculated by incorporating the static energy for (a) X, X$'$ = Ti, Zr, Hf; Y= Ni; Z= Sn, (b) X= Zr; Y, Y$'$ = Ni, Pd, Pt; Z= Sn, and (c) X = Zr; Y= Ni; Z= Ge, Sn, Pb.}
\label{fig:FE}
\end{figure}

The formation energy~($E_{\mathrm{f}}$) of the compositions is calculated for all pristine compounds and their substituted counterparts using eqn.~\ref{Eform}. The formation energy of the 50\% substituted compounds were found to be generally in between the formation energy of their pristine parents with only one exception of the Y site substitution with ZrNi$_{0.5}$Pt$_{0.5}$Sn, whereby the formation energy of the substituted compound is found to lower than both its parent counterpart. However, all the half-Heusler parent and the substituted compounds for the above mentioned compositions and stoichiometries are found to have negative formation energy, as shown in Fig.~\ref{fig:FE}, which indicates that all these structures can be realised in the laboratory. Furthermore, the formation energy of the parents and the 50 \% substituted ones are also verfied to be negative by taking into account of the thermodynamic free energy of the corresonding structures at 300 K, which has been provided in the section S1 of the supplementary information.  

\subsection{Electronic structure analysis}
\begin{figure} 
\centering
\includegraphics[scale=0.3]{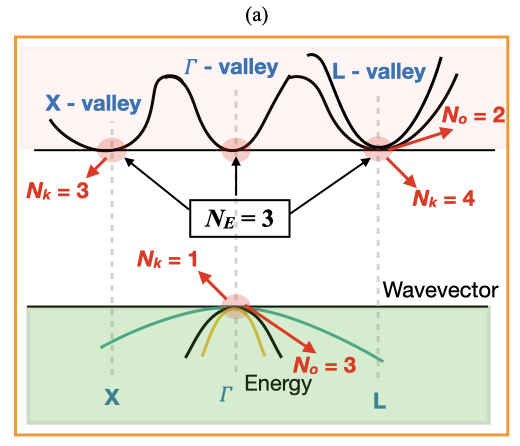}
\centerline{\includegraphics[scale=.4]{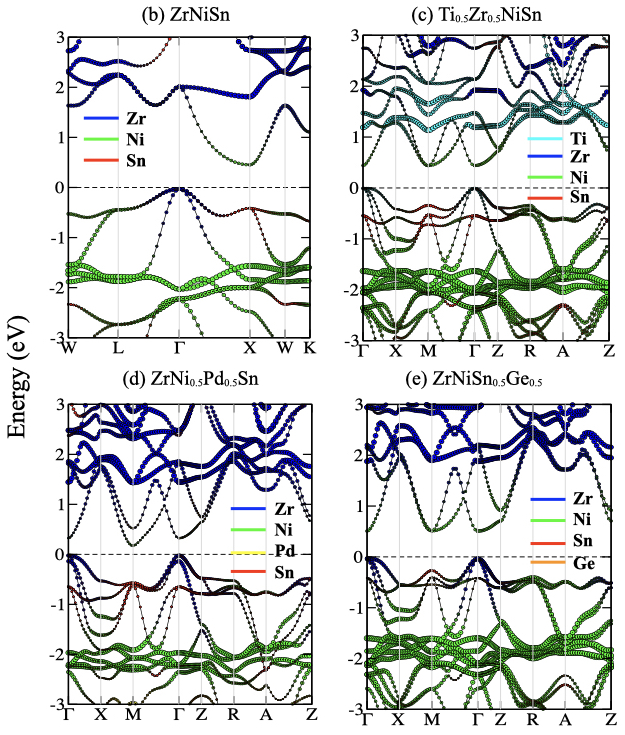}}
\caption{(a) Schematic representation of the different degeneracies in the electronic band structure of a symmetric solid. The electronic band structure (GGA-PBE) of (b) parent ZrNiSn and a few respective cases of X, Y and Z site substitution viz., (c) Ti$_{0.5}$Zr$_{0.5}$NiSn, (d) ZrNi$_{0.5}$Pd$_{0.5}$Sn, and (e) ZrNiGe$_{0.5}$Sn$_{0.5}$.}
\label{fig:band}
\end{figure}

The electronic band structure~(eBS) of the compounds may be analysed in terms of the several factors as has been enumerated below (as demonstrated in the schematics of Fig.~\ref{fig:band}(a));
\begin{itemize}
\item Orbital degeneracy, N$_\mathrm{o}$, which is the number of bands converging at the VBM or CBm at the same K point).
\item K point degeneracy, N$_\mathrm{k}$, which is the degeneracy of equivalent valleys due to the high symmetry point in the Brillouin zone (i.e. N$_\mathrm{k}$ = 1 for $\Gamma$ point and N$_\mathrm{K}$ = 3 for the $X$ point).
\item Valley/tip degeneracy N$_\mathrm{E}$, which is the number of tips lying at the same energy level for all different K points.
\item Band/total degeneracy, N$_\mathrm{v}$, which is derived from all these degeneracies collectively for a given high symmetry point.
\end{itemize} 

Using these notations, the eBS of the parent as well as the substituted compounds are analysed in the following. 

\subsubsection{Parent half Heusler compounds}
\label{sec:single}

The primitive cell of the XYZ parent compounds, containing three atoms (one of each species) per unit cell, is considered for performing the eBS calculation. The eBS is plotted along high symmetry path W-L-$\Gamma$-X-W-K of the irreducible first Brillouin zone (as shown in Fig.~\ref{fig:band}~(b)). All the parent compounds show similar trends in terms of band degeneracies. For instance, the VBM at $\Gamma$ point ($\mathrm{N_k} =1$) has three fold orbital degeneracy ($\mathrm{N_o} =3$) leading to the band degeneracy of $\mathrm{N_v} =3$, while the CBm has a single valley at the X point ($\mathrm{N_k} =3$) leading the band degeneracy to $\mathrm{N_v}=3$). All the selected parent compounds (i.e. TiNiSn, ZrNiSn, HfNiSn, ZrPdSn, ZrPtSn, ZrNiGe, ZrNiPb) are found to be indirect semiconductor, with their band gap varying in the range of $\sim$ 0.35 - 1.0 eV. Although majority of the bandgap values show fair agreements with other theoretical reports~\cite{Uher1999,Zou2013,ZFeng2019}, which are approximately $\sim$ 0.3 eV higher than the corresponding experimental results (0.12, 0.19 and 0.22 eV for TiNiSn, ZrNiSn, and HfNiSn, respectively)~\cite{Aliev1989, Aliev1987, YYan2018}. This trend is anomalous to this specific class of compound, since DFT in the GGA level generally underestimates the band gap of semiconductors. The reason for this inconsistency has been attributed to the presence of structural imperfections, such as defects, boundaries etc in the experimental samples. 

The effect of spin orbit coupling~(soc) on the eBS is also explored for these compounds, which is however, found to have very minor effect~(see supplimentary information for details). Thus, as revealed from the preliminary band structure analysis, no particular parent XYZ compound is found to be superior or inferior to the other as per as its thermoelectric application is concerned, since they all have narrow band gap with same features in the eBS. Therefore, for the X site substitution, all the three elements viz.~Ti, Zr, and Hf are selected for isoelectronic substitution, while the elemental composition for Y~(=Ni) and Z~(=Sn) site is kept fixed keeping the non toxicity and earth abundancy in account. It is to be noted here that the selected compositions viz. TiNiSn, ZrNiSn and HfNiSn are also the most studied n-type Heusler compounds. Thus, the compositions with X$_{p}$X$'_{1-p}$NiSn, with X, X$'$= Ti, Zr, Hf, are explored for the X site substitutions as discussed in the succeeding section.\\

\subsubsection{X site substitution}
\label{sec:X}

The eBS of X$_{p}$X$'_{1-p}$NiSn compositions with X, X$'$= Ti, Zr, Hf is analysed for different stoichiometries viz. $p= 0.25, 0.5, 0.75$. As discussed earlier, the symmetry of the structure of all compositions with $p = 0.25$ and $p = 0.75$ i.e. X$_{0.25}$X$'_{0.75}$NiSn and X$_{0.75}$X$'_{0.25}$NiSn is found to be identical to the conventional parent lattice. The corresponding band structure of the representative compounds is included in the supplimentary information. These substitutions result in a direct band gap semiconductor, with the CBm lying above the VBM at the $\Gamma$ point. The $p =0.5$ case, on the other hand, has increased number of valley and band degeneracies as compared to the parent as well as the substituted ones with $p= 0.25, 0.75$ stoichiometries for all compositional variation. As has been discussed in details in the supporting information, the electronic transport properties of the $p=0.5$ case is only marginally higher than those of $p= 0.25, 0.75$. However, major favorable variation in $\kappa_{l}$ is expected for $p=0.5$ as compared to $p= 0.25, 0.75$ empirically as is also verfied by our DFT results (as explained in details in the section S5.1 of the supporting information). Thus, the eBS of the $p=0.5$ is analysed hereby in details. 

For $p =0.5$, all compositions are found to have a direct band gap at the $\Gamma$ point in the range of $\sim$ 0.40 - 0.45 eV, with another energy degenerate band pocket forming at the M point, where two bands merge~(see Fig.~\ref{fig:band}(c)). The energy eigenvalues of bands and sub bands of substituted compounds corresponding to the electronic band structure (Fig.~\ref{fig:band}(c)) are tabulated in table.~S2 and ~S3 of the supplementary information. As it can be seen that the energy eigenvalues of sub-bands at M points of Ti$_{0.5}$Zr$_{0.5}$NiSn are same, whereas energy difference between degenerate bands (conduction band edges, CBE) M and $\Gamma$ is very small i.e., 0.002 eV. Whereas, the VBM retains three fold orbital degeneracy at the $\Gamma$ point. This means that both the CBEs simultaneously participate in transport and may improve the thermoelectric performance of the compound. Therefore, it may be undoubtedly inferred that all the X$_{0.5}$X$'_{0.5}$NiSn compositions will serve as excellent basis for further $n$-type or $p$-type doping. However, in order to select one best composition  Ti$_{0.5}$Zr$_{0.5}$NiSn is chosen, because of the earth abundance of Ti over Hf. The $\kappa_{l}$ of all the X site substituted compounds are also compared eventually, which revealed that this choice has comparable $\kappa_{l}$ with the other X site substituted compositions.  

\subsubsection{Y site substitution}
\label{sec:Y}

For the Y site substitution, the eBS of the compositions Ti$_{0.5}$Zr$_{0.5}$Y$_q$Y$'_{1-q}$Sn for Y, Y$'$ = Ni, Pd, Pt is explored for $q =0.25, 0.5, 0.75$. However, in order to highlight the adverse effect of Y site substitution alone, the eBS of the composition ZrNi$_{0.5}$Pd$_{0.5}$Sn~(as shown in Fig.~\ref{fig:band}(d)), is presented and analysed as the representative case (while the other stoichiometries are included in the supplementary information in section S3.5). From this, several negative effects of Y site substitution in the eBS is identified, viz. the splitting of the orbital and valley degeneracy collectively leading to lowering in the band degeneracies drastically at the CBm and VBM. For instance, in ZrNi$_{0.5}$Pd$_{0.5}$Sn (~Fig.~\ref{fig:band}(d), the CBm is found at M point ($\mathrm{N_o =1}$), while the VBM at $\Gamma$ point ($\mathrm{N_o =2}$). An energy difference of $\sim$ 0.15 eV is observed between the CBm at the $\Gamma$ and M point, while energy difference of $\sim$0.35 eV is observed between the two conduction band sub-bands at the M point. Also, energy difference of $\sim$ 0.12 eV is observed between the hole bands at the $\Gamma$-point of VBM~(see supplementary information, see table S2 and S3). The lowering of the bands at the M point also results in reduction of band gap in all cases (in the range of $\sim$ 0.15 - 0.22 eV) and all compositions are found to be indirect semiconductor. Thus, it can be inferred that the Y site substitution in HH is not particularly advantageous for thermoelectric application. Moreover, the lowering of the band gap in these compositions may also give rise to adverse bipolar effects in the transport coefficients, due to which both charge carriers may cancel each other even at room temperature. 

In order to explore the reason behind the splitting of the degenerate bands at the M point, which is observed uniquely only for the Y site substitution, the local symmetry around the substituted atom in the tetragonal structure is investigated. As for instance, the local symmetry surrounding the Pd atom at the center of the tetragonal ZrNi$_{0.5}$Pd$_{0.5}$Sn crystal is found to be broken because of the pesence of two different atoms as the first nearest neighbor (leading to different bond lengths arising from the Pd-Sn and the Pd-Zr). However, such a situation does not arise for X site or Z site substitution, where the substituted atom sitting at the center always has the same atom as its first nearest neighbor and for its second nearest neighbor. Therefore, the local symmetry is retained for the X site as well as the Z site substitution~(see supplementary Fig. S11 and the discussion there in for more details).         

Thus, the Ti$_{0.5}$Zr$_{0.5}$NiSn is still retained as the best starting composition for the Z site substitution, which is discussed in the subsequent section. 

\begin{figure}
\centering
\includegraphics[scale=0.41]{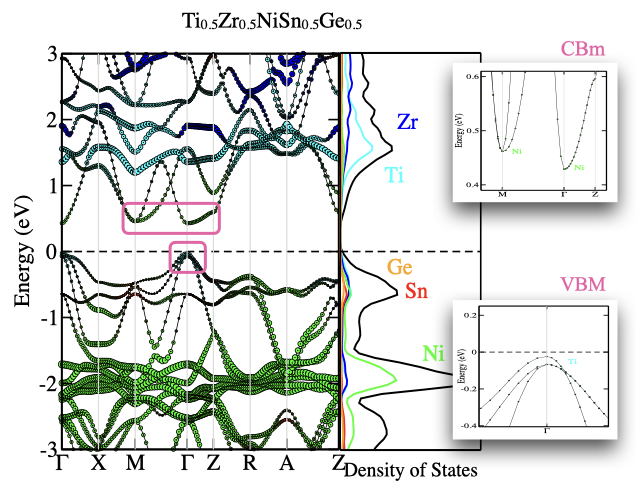}
\caption{Electronic bandstructure and projected partial density of states of Ti$_{0.5}$Zr$_{0.5}$NiGe$_{0.5}$Sn$_{0.5}$. Insets highlight the region near the VBM as well as the CBm.}
\label{fig:XZ}
\end{figure}

\subsubsection{Z site substitution}
\label{sec:Z}

\begin{figure}
\centering
\centerline{\includegraphics[scale=0.42]{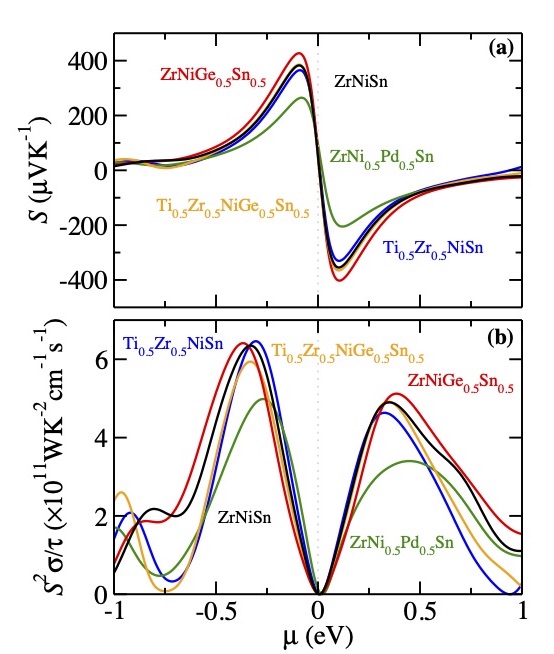}}
\caption{Seebeck coefficient ($S$) and power factor ($S^{2}\sigma/\tau$) plotted as a function of electronic chemical potential $\mu$ for few selected substituted compounds.}
\label{fig:S}
\end{figure}

For the Z site substitution, the eBS of the compositions Ti$_{0.5}$Zr$_{0.5}$NiZ$_r$Z$'_{1-r}$ for Z, Z$'$ = Ge, Sn, Sb is analysed for $r =0.25, 0.5, 0.75$. However, in order to highlight the sole effect of Z site substitution, the eBS of the composition ZrNiGe$_{0.5}$Sn$_{0.5}$~(as shown in Fig.~\ref{fig:band}(e)), is also presented as the representative case (while the other stoichiometries are included in the supplementary information). For all compositions with $r =0.5$, the Z site substitution, however, shows exactly the same trend as of X site substitution in terms of band and valley degeneracies, with band gap lying in range $\sim$ 0.54 - 0.60 eV. For instance, in Z site substituted ZrNiSn$_{0.5}$Ge$_{0.5}$ compound, same energy value is observed at sub bands of M point and a very small energy difference of $\sim$0.005 eV is observed between the two CBEs (M and $\Gamma$ point) (see supplementary information; table S2 and S3).

The eBS of Ti$_{0.5}$Zr$_{0.5}$NiGe$_{0.5}$Sn$_{0.5}$ is plotted in Fig.~\ref{fig:XZ} as a representative case of co site doping. The orbital degeneracy at the M point ($\mathrm{N_o} =2$) and valley degeneracy ($\mathrm{N_E} =2$), with energy difference of $\sim$0.03 eV between sub bands of the CBE is retained while one band in the VBM at the $\Gamma$ point undergo splitting ($\Delta E \sim0.04$ eV) as shown in the insets of Fig.~\ref{fig:XZ}. This may lower the power factor in case of $p$-type doping of the material. However, this does not affect the $n$-type doping scenario. The co-site doping, however, provides the possibility of tuning the $\kappa_{l}$ via introduction of mass disorder~(see section \ref{sec:trans} for further discussion). 

Overall, the X, Z and X-Z co-site substitution lead to an increase in band degeneracy in the CBm, which offers additional band effective masses with corresponding band degeneracies ($\mathrm{N_v}$) at the high symmetry points. High degeneracy $\mathrm{N_v}$ within the degenerate energy window of transport can improve the Seebeck coefficient while retaining the high mobility by maintaining low scattering. All these effects may collectively enhance the power factor significantly. But in order to inspect the combined scenario, the transport properties of the representative cases are calculated using the semi classical Boltzmann transport equation.      

\begin{table}[h]
\begingroup
\caption{\label{tab:3}The transport properties of few representative viz. parent (ZrNiSn), X-site (Ti$_{0.5}$Zr$_{0.5}$NiSn), Y-site (ZrNi$_{0.5}$Pd$_{0.5}$Sn), Z-site (ZrNiGe$_{0.5}$Sn$_{0.5}$) and co-site (Ti$_{0.5}$Zr$_{0.5}$NiGe$_{0.5}$Sn$_{0.5}$) substituted compounds. The Seebeck $S$~($\mu$VK$^{-1}$), relaxation time scaled electrical conductivity $\frac{\sigma}{\tau}$ (10$^{20}\times \Omega^{-1}$m$^{-1}$s$^{-1}$), relaxation time scaled power factor $\frac{S^{2}\sigma}{\tau}$~(10$^{11}\times$WK$^{-2}$m$^{-1}$s$^{-1}$), the electronic ($zT_\mathrm{el}$) and total figure of merit ($zT_\mathrm{total}$), for the electron chemical potential $\mu$ (eV) where their maximum value is attained for the $n$-type and $p$-type doping at 700 K.}
\renewcommand{\arraystretch}{1.1}
\begin{tabular}{cccccccc}
\hline
System&\hspace*{0.00mm}type & $\mu$ &$S$& $\frac{\sigma}{\tau}$ & $\frac{S^{2}\sigma}{\tau}$& $zT_\mathrm{el}$ & $zT_\mathrm{total}$ \\
\hline
ZrNiSn & $n$ & 0.35& -141 &  0.25 & 4.90 & 0.50&0.35\\
          & $p$ &-0.33& 155 & 0.26 & 6.35 & 0.57&0.42\\ 
X-site & $n$ & 0.32& -146 &  0.22 & 4.63 &0.54&0.45\\
        &      $p$      &-0.30& 162 & 0.25 & 6.46 &0.61&0.52\\ 
Y-site & $n$ &0.45& -87& 0.45 & 3.40 &0.25&0.12\\
          &     $p$     &-0.27 &139 & 0.26  & 4.98 &0.48&0.20\\
Z-site & $n$ &0.38 & -143& 0.25 & 5.12 &0.53&0.46\\
          &     $p$     &-0.36 &156& 0.26 & 6.41&0.59&0.52\\
Co-site & $n$ &0.34& -150& 0.22 & 4.92 &0.56&0.49\\
         &     $p$     &-0.33 &157& 0.24  & 5.94&0.59&0.53\\
\hline
\end{tabular} 
\endgroup
\end{table}

\subsection{Transport properties}
\label{sec:trans}

\subsubsection{Electronic transport properties}

Finally, the transport properties of all the filtered compositions viz. X$_{0.5}$X$'_{0.5}$NiSn and X$_{0.5}$X$'_{0.5}$NiZ$_{0.5}$Z$'_{0.5}$ for  X, X$'$ = Ti, Zr, Hf and Z, Z$'$ = Ge, Sn are calculated. Few of the representative ones are hereby discussed. The Seebeck coefficient~($S$)and power factor scaled by the relaxation time ($\frac{S^2\sigma}{\tau}$) are plotted as a function of chemical potential ($\mu$) of electrons and holes in Fig.~\ref{fig:S} (a) and (b) respectively, while the maximum values attained and the corresponding chemical potential are given in table.~\ref{tab:3}. The positive and negative value of the chemical potential ($\mu$) indicate that the dopants are electrons ($n$-type) and holes ($p$-type), respectively. Please note that the scenario of the $n$-type (positive $\mu$) as well as $p$-type doping (negative $\mu$) can be inferred from the given plot itself. However, in reality in order to achieve such $\mu$ in laboratory, one has to dope the compound with small $n$ or $p$-type impurities, which may introduce aditional disorder and effectively tune the transport coefficients as well. The $S$ and hence the $\frac{S^2\sigma}{\tau}$ for the X, Z and co site substituted compounds are found to be comparable with the parent composition. The shape of the lowest conduction band which is contributing to transport is almost the same for Ti$_{0.5}$Zr$_{0.5}$NiSn, ZrNiGe$_{0.5}$Sn$_{0.5}$ and Ti$_{0.5}$Zr$_{0.5}$NiGe$_{0.5}$Sn$_{0.5}$, resulting in similar effective masses (see table.~\ref{tab:mass}). Because of many fold orbital and valley degeneracies, $n$-type doping may yield higher power factor. Most importantly, the Z site substituted composition ZrNiGe$_{0.5}$Sn$_{0.5}$ is found to have the best power factor for both $n$ as well as $p$-type doping, which was hitherto not explored experimentally for this particular stoichiometry. The reason for the high power factor of ZrNiGe$_{0.5}$Sn$_{0.5}$ is the higher curvature of all the degenerate bands (at CBm) as compared to the X-site substituted Ti$_{0.5}$Zr$_{0.5}$NiSn and co-site substituted Ti$_{0.5}$Zr$_{0.5}$NiGe$_{0.5}$Sn$_{0.5}$ as tabulated in table \ref{tab:mass}. Because of the decrease in band degeneracies for the Y site substitution, it results in lowest $S$ and $\frac{S^2\sigma}{\tau}$ as compared to all other substituted counterparts. Furthermore, the anisotropy in the transport coefficients is analysed for the 50 \% subsituted compositions (with tetragonal geometry) and no strong anisotropy is
observed in any of the reported compounds along different directions (see supplementary information SI. S5.1.1 for corresponding discussion).

\begin{table}
\begingroup
\caption{\label{tab:mass}Effective mass ($m^{*}/m_{e}$) of few selected compounds i.e. the parent (ZrNiSn) and the X-site (Ti$_{0.5}$Zr$_{0.5}$NiSn), Y-site (ZrNi$_{0.5}$Pd$_{0.5}$Sn), Z-site (ZrNiGe$_{0.5}$Sn$_{0.5}$) and co-site (Ti$_{0.5}$Zr$_{0.5}$NiGe$_{0.5}$Sn$_{0.5}$) substituted compounds for the different bands meeting at the VBM (indexed as VBM, VBM-1 and VBM-2 respectively) and CBm (indexed as CBm and CBm+1 respectively) along the different high symmetry paths in Brillouin zone (Dir).}
\setlength{\tabcolsep}{3pt}
\begin{tabular}{p{1cm}p{1cm}p{1cm}p{1cm}p{1cm}p{1cm}p{1cm}p{1cm}}
\hline
Dir. &Band Index & \multicolumn{5}{c}{Effective mass} \\
\hline
 & & ZrNiSn & X-site & Y-site & Z-site & co-site \\
\hline
$\Gamma	\rightarrow \mathrm{L}$& VBM  &2.5\\
$\Gamma	\rightarrow \mathrm{X}$&   &0.9\\
$\Gamma	\rightarrow \mathrm{L}$&VBM-1&  0.3\\
$\Gamma	\rightarrow \mathrm{X}$& &  0.4 \\

$\Gamma \rightarrow \mathrm{M}$& VBM &&  1.1 & 0.5 &0.6 &1.4\\
$\Gamma \rightarrow \mathrm{Z}$&  &&  0.5 & 0.8 &0.8 &0.5\\

$\Gamma \rightarrow \mathrm{M}$&VBM-1& & 1.1 & 0.4 &0.4 &0.8 \\
$\Gamma \rightarrow \mathrm{Z}$& & & 1.1 & 0.8 &0.8 &1.1 \\
$\Gamma \rightarrow \mathrm{M}$&VBM-2& & 0.5 & &0.8&0.5\\
$\Gamma \rightarrow \mathrm{Z}$& & & 1.1 & &0.4&1.1\\
X$ \rightarrow \mathrm{\Gamma}$ &CBm&  3.2 &   & \\
X$ \rightarrow \mathrm{W}$ &&  0.4 &   & \\
$\Gamma \rightarrow \mathrm{M}$ &&   & 0.5 & &0.4 &0.5 \\
$\Gamma \rightarrow \mathrm{Z}$ &&   & 3.3 & &6.9 &5.1\\
M$ \rightarrow \mathrm{X}$ & &   & 0.9 &0.6&1.0&1.1 \\
M$ \rightarrow \mathrm{\Gamma}$ &&   & 3.3 &0.6&6.6&7.5\\
M$ \rightarrow \mathrm{X}$ &CBm+1&  & 0.7 & &0.6&0.8\\
M$ \rightarrow \mathrm{\Gamma}$& &  & 0.4 & &0.4&0.5\\
\hline
\end{tabular}
\endgroup
\end{table}

\subsubsection{Lattice thermal conductivity}

\begin{figure}
\centering
\includegraphics[scale=.27]{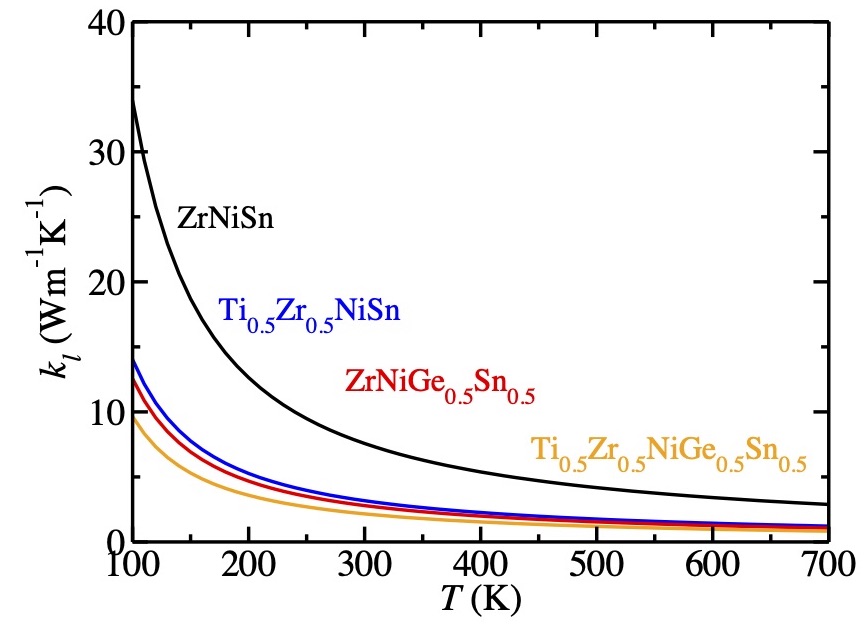}
\caption{Lattice thermal conductivity ($\kappa_{l}$) of few selected compounds plotted as function of temperature $T$, calculated using Asen Palmer modified version of Debye Callaway theory.}
\label{fig:kl}
\end{figure}

The Asen-Palmer modified version of the Debye Callaway theory parametrized for solid~(see section S4 and S5 of supplementary information for details) is used to model the phonon lifetimes for the normal as well as the Umklapp scattering processes by including the contributions stemming from the transverse acoustic (TA and TA$'$) and the longitudinal acoustic (LA) phonon modes (from their respective mode group velocities ($v_i$), Gr$\ddot{\mathrm{u}}$neisen parameter ($\gamma_{i}$) and Debye temperatures ($\mathrm{\theta}_{i}$) for $i$ =TA, TA$'$, LA modes. The corresponding plots for the phonon spectrum, group velocity and Gr$\ddot{\mathrm{u}}$neisen parameter is included as supplementary material. The corresponding lattice thermal conductivity ($\kappa_{l}= \sum \kappa_{i}$, for $i$ =TA, TA$'$, LA) is calculated from the corresponding values of the $v_i$, $\gamma_{i}$, $\mathrm{\theta}_{i}$ etc, which are given in table~\ref{tab:kl}. The $\kappa_{l}$ of the compounds with high power factors viz. the parent (ZrNiSn), X-site (Ti$_{0.5}$Zr$_{0.5}$NiSn), Z-site (ZrNiGe$_{0.5}$Sn$_{0.5}$) and co-site (Ti$_{0.5}$Zr$_{0.5}$NiGe$_{0.5}$Sn$_{0.5}$) substituted compounds are only presented for the sake of brevity~(see Fig.~\ref{fig:kl} and table~\ref{tab:kl}). The $v_i$s and average group velocity of sound ($v_\mathrm{s}$) are found to be similar for all the selected compounds. The average Debye temperature $\theta_{D}$, is found to be the lowest for X site (Ti$_{0.5}$Zr$_{0.5}$NiSn) while the highest for the co-site (Ti$_{0.5}$Zr$_{0.5}$NiGe$_{0.5}$Sn$_{0.5}$) substituted case. It is to be noted that the integration for $\kappa_{l}$ is limited by the $\mathrm{\theta}_{i}$s and hence lower $\mathrm{\theta}_{i}$s indicate lower $\kappa_{l}$. However, the $\gamma_{i}$s are found to be different for the compounds, which play the deterministic factor in the change observed in the $\kappa_{l}$. The $\gamma_{i}$s are found to be the lowest for ZrNiSn, comparable for Ti$_{0.5}$Zr$_{0.5}$NiSn and ZrNiGe$_{0.5}$Sn$_{0.5}$ while the highest for Ti$_{0.5}$Zr$_{0.5}$NiGe$_{0.5}$Sn$_{0.5}$, pertaining to the highest mass disorder in similar cell volume. The high anharmonicity, as indicated by the high $\gamma_{i}$s in Ti$_{0.5}$Zr$_{0.5}$NiGe$_{0.5}$Sn$_{0.5}$, is responsible for the increased scattering centers and hence, lowest $\kappa_{l}$ in the compound. In summary, the $\kappa_{l}$ of the substituted compounds is found to be significantly lower than the parent compound (by more than 60 \%). However, in this formalism, the effect of defect phonon scattering is not incorporated explicitly, which arises from the mass and strain fluctuations stemming from the substitutions~\cite{Callaway1959,JYang2004}. If explicitly taken into account it may still reduce the $\kappa_{l}$ for the 50 \% substituted ones as compared to the 25 \% and 75 \% substitutions at a given site (see supplimentary information SI S5.5 for corresponding discussion).

\subsubsection{Figure of merit}
The dimensionless figure of merit of a thermoelectric material ($zT$) can be written as the sum of the electronic and lattice figure of merit ($\frac{1}{zT}$ = $\frac{1}{zT_{el}}$ + $\frac{1}{zT_{l}}$). $zT_{el}$ is calculated from transport parameters obtained in BoltzTrap viz. S, $\frac{\sigma}{\tau}$, $\frac{\kappa_{el}}{\tau}$ as $zT_{el}=\frac{S^{2}\sigma/\tau T}{\kappa _{e}/\tau}$. Whereas, $zT_{l}$ is calculated as $zT_{l}=\frac{S^{2}\sigma/\tau T}{\kappa _{l}}$ (where $\tau$ is choosen to be 10$^{-14}$ s)), and $\kappa_{l}$ is calculated using the Debye Callaway formalism as discussed before. The $zT$ for X-, Y-, Z- and co-substituted ZrNiSn is also given in table.~\ref{tab:3}. Pertaining to its lowest power factor, the $zT$ of the Y-site subsituted ZrNi$_{0.5}$Pd$_{0.5}$Sn is found to be the lowest, which show a reduction of 65 \% ($n$-type) and 52\% ($p$-type) in the $zT$ as compared to the parent compound. The $zT$ of X-site substituted Ti$_{0.5}$Zr$_{0.5}$NiSn and Z-site substituted ZrNiGe$_{0.5}$Sn$_{0.5}$ is found to comparable for both $p$- as well as $n$- type doping. This is found to be approximately 24~\% (for $n$-type) and 20~\% (for $p$-type) higher than the parent ZrNiSn, which stems from the low $\kappa_{l}$ of the substituted compounds. Pertaining to its lowest $\kappa_{l}$, the co-site substituted Ti$_{0.5}$Zr$_{0.5}$NiGe$_{0.5}$Sn$_{0.5}$ shows comparable $zT$ as X and Z site substituted compounds, inspite of comparatively lower power factor for the $p$-type doping. 

\begin{table}
\caption{\label{tab:kl}
The mode resolved~($v_i$) and average~($v_\mathrm{s}$)~phonon group velocity~(Km/s), mode resolved Gr$\ddot{\mathrm{u}}$neisen parameter~($\gamma_i$), mode resolved~($\mathrm{\theta}_{i}$) and average ($\mathrm{\theta}_\mathrm{D}$) Debye temperature~(K) [$\mathrm{\theta}_\mathrm{D}$ = $\frac{\hbar}{k_{B}}\left ( \frac{3N}{4\pi V} \right )^{1/3}\upsilon _{s}$] of the transverse acoustic~($i$ = TA/TA$'$) and longitudinal acoustic~($i$ = LA) branches. $v_i$ and $\gamma_i$ are calculated from the maximum group velocity and Gr$\ddot{\mathrm{u}}$neisen parameter attained by the phonon modes near the $\Gamma$ point. $v_\mathrm{s}$ is calculated from $v_\mathrm{TA}$, $v_\mathrm{TA'}$, and $v_\mathrm{LA}$ using the expression $v_\mathrm{s}= [\frac{1}{3}(\frac{1}{{v_\mathrm{LA}}^3} +\frac{1}{{v_\mathrm{TA}}^3} +\frac{1}{{v_\mathrm{TA'}}^3})]^\frac{-1}{3}$. Finally, the $\kappa_\mathrm{l}$ is calculated using the Debye Callaway formalism and the corresponding value at 300 K is enlisted for few selective compunds viz. the parent (ZrNiSn), X-site (Ti$_{0.5}$Zr$_{0.5}$NiSn), Z-site (ZrNiGe$_{0.5}$Sn$_{0.5}$) and co-site (Ti$_{0.5}$Zr$_{0.5}$NiGe$_{0.5}$Sn$_{0.5}$) substituted compounds.}
 \begin{tabular*}{0.48\textwidth}{@{\extracolsep{\fill}}lcccccccccc} 
 \hline
System & Modes ($i$) & $v_i$  & $v_\mathrm{s}$ & $\gamma_i$ & $\theta_i$ & $\theta_\mathrm{D}$  &  $\mathrm{\kappa}_\mathrm{l}\mid_{300}$ \\     
\hline
               & TA           &2.6    &          & 1.5   & 135 &           &  &\\
ZrNiSn   & TA$'$            &2.7   & 3.0     &1.6    & 140&    332 & 7.6 &\\
              & LA       &4.6   &           &1.8    &148&            &       &\\
\hline
        & TA & 2.6 &     &1.8  &  111 &     &     & \\
X-site  & TA$'$ &  2.7 & 2.9 & 2.1  &  112 & 326 & 3.2 &\\
       & LA &  4.3 &     & 2.2  & 138 &     &       &\\
\hline                 
      & TA  &   2.7 &     & 2.1 & 120 &      &   & \\
Z-site  & TA$'$  & 2.7 & 3.0 &  2.1 & 121 &  339 &  2.8 &\\
      & LA   & 4.4 &     & 2.2 & 146 &      &    &\\
\hline
        & TA &  2.7 &     & 2.2 &  120 &      &     &\\                 
Co-site & TA$'$  & 2.7 & 3.0 & 2.2 & 122 & 389 &  2.2 &\\
        & LA &  4.6 &     & 3.5 & 154 &      &    & \\
\hline        
\end{tabular*}
\end{table}

\section{Conclusions}
The structural, electronic and transport properties of 18 valence electron count based X$_{p}$X$'_{1-p}$Y$_{q}$Y$'_{1-q}$Z$_{r}$Z$'_{1-r}$ (where X, X$'$= Ti, Zr, Hf; Y, Y$'$ = Ni, Pd, Pt and Z, Z$'$ = Ge, Sn, Pb and $p$, $q$, and $r$ = 0, 0.25, 0.5, 0.75, 1) half-Heusler compounds is investigated using the high throughput first-principles density functional theory calculations. An intelligent scanning strategy is adapted. As the first step, the electronic structure of all the parent compounds is explored in terms of parameters such as the band degeneracy, valley degeneracy, band gap etc, which are generally the most important parameters for improving the power factor. The parent compounds viz. TiNiSn, ZrNiSn and HfNiSn are selected, whereby the Y and Z elements are kept fixed because of their non-toxicity and earth abundance. Subsequently, the elemental composition at the X site is varied as X$_{p}$X$'_{1-p}$NiSn with X, X$'$= Ti, Zr, Hf for different stoichiometries viz. $p= 0.25, 0.5, 0.75$. All the substituted compounds are found to have negative formation energy, which is calculated by the static energy calculations. From the band engineering perspective, all these combinations are found to have similar electronic structure as long as the stoichiometries are retained. However, the among the different stoichiometries, the X$_{0.5}$X$'_{0.5}$NiSn are found to have higher band and valley degeneracies, indicating improvement in power factor. Thus, Ti$_{0.5}$Zr$_{0.5}$NiSn is chosen as the representative case from the X site substituted compositions (because of the earth abundance of the elements) for proceeding to subsequent substitutions. As the next step, the electronic structure of the Y site substituted compositions viz. Ti$_{0.5}$Zr$_{0.5}$Y$_q$Y$'_{1-q}$Sn for Y, Y$'$ = Ni, Pd, Pt is explored for $q =0.25, 0.5, 0.75$. However, all the Y site substitutions led to breaking of the degeneracy of the bands and hence, are not found to be effective for improvising the electronic power factor of the compounds. Subsequently, the electronic structure of the Z site substituted compositions viz. Ti$_{0.5}$Zr$_{0.5}$NiZ$_r$Z$'_{1-r}$ for Z, Z$'$ = Ge, Sn, Sb with $r =0.25, 0.5, 0.75$ is explored. Our results reveal that the X site and Z site doping may be independently opted for improving the power factor, while the X and Z co-site doping may lead to lowering in power factor for the p-type material. As the last step, the lattice thermal conductivity of these representative compounds is calculated using the Asen Palmer modified version of Debye Callaway theory parameterized for solid. The substituted compounds are found to have $\mathrm{\kappa}_\mathrm{l}$ reduced by atleast $\sim$ 60 \%, which is due to the intoduction of mass disorder in the same lattice volume. Finally, the $zT$ of X-site substituted Ti$_{0.5}$Zr$_{0.5}$NiSn and Z-site substituted ZrNiGe$_{0.5}$Sn$_{0.5}$ (being comparable for both $p$- as well as $n$- type doping) is found to be approximately 24~\% (for $n$-type) and 20~\% (for $p$-type) higher than the parent ZrNiSn, which stems from the low $\kappa_{l}$ of the substituted compounds. However, the co-site substituted composition (Ti$_{0.5}$Zr$_{0.5}$NiGe$_{0.5}$Sn$_{0.5}$), is found to have comparable $zT$ for both the $p$- as well as $n$- type doping, which is preferential for its device application. 

In summary, DFT calculations are performed to explore the thermoelectric properties of the X, Y and Z substituted half Heusler compounds without restricting the compositions to any prior biases or conclusions from the previous literature. The calculations confirm several known facts as well as reveal several unknown ones. Exemplarily, the Y site substitution is generally avoided experimentally (if at all, there are only few reports of very minute doping at the Y site, which has improved the $zT$). This study confirms the scenario and reveals the predicament behind such substitution. However, the Z site substitution, leads to enhancement of the powerfactor, which is comparable to the scenario of X site substitution. However, the Z site substitution has not been hitherto explored thoroughly, which may provide further scope in tuning the $zT$ of these compounds. Thus, this study may serve as a consolidated blue print for experimentalist working with similar compositions for enhancing the powerfactor and $zT$ of the compositions.

\section*{Acknowledgements}
The authors thank Prof. Titas Dasgupta for many fruitful discussions. AB acknowledges the DST Inspire faculty project (DST/INSPIRE/04/2015/000089), IIT B seed grant project (RD/0517-IRCCSH0-043) and SERB ECRA project (ECR/2018/002356) for the financial assistance. The high performance computational facilities (viz. dendrite, spacetime, and corona) of IIT Bombay and of CDAC (Param Yuva-II) are acknowledged for providing the computation hours.

%%%END OF MAIN TEXT%%%


\balance


%%%REFERENCES%%%
\bibliography{rsc} %You need to replace "rsc" on this line with the name of your .bib file
\bibliographystyle{rsc} %the RSC's .bst file

\end{document}

\pagestyle{fancy}
\thispagestyle{plain}
\fancypagestyle{plain}{

%%%HEADER%%%
\renewcommand{\headrulewidth}{0pt}
}
%%%END OF HEADER%%%

%%%PAGE SETUP - Please do not change any commands within this section%%%
\makeFNbottom
\makeatletter
\renewcommand\LARGE{\@setfontsize\LARGE{15pt}{17}}
\renewcommand\Large{\@setfontsize\Large{12pt}{14}}
\renewcommand\large{\@setfontsize\large{10pt}{12}}
\renewcommand\footnotesize{\@setfontsize\footnotesize{7pt}{10}}
\makeatother

\renewcommand{\thefootnote}{\fnsymbol{footnote}}
\renewcommand\footnoterule{\vspace*{1pt}% 
\color{cream}\hrule width 3.5in height 0.4pt \color{black}\vspace*{5pt}} 
\setcounter{secnumdepth}{5}

\makeatletter 
\renewcommand\@biblabel[1]{#1}            
\renewcommand\@makefntext[1]% 
{\noindent\makebox[0pt][r]{\@thefnmark\,}#1}
\makeatother 
\renewcommand{\figurename}{\small{Fig.}~}
\sectionfont{\sffamily\Large}
\subsectionfont{\normalsize}
\subsubsectionfont{\bf}
\setstretch{1.125} %In particular, please do not alter this line.
\setlength{\skip\footins}{0.8cm}
\setlength{\footnotesep}{0.25cm}
\setlength{\jot}{10pt}
\titlespacing*{\section}{0pt}{4pt}{4pt}
\titlespacing*{\subsection}{0pt}{15pt}{1pt}
%%%END OF PAGE SETUP%%%

%%%FOOTER%%%
\fancyfoot{}
\fancyfoot[LO,RE]{\vspace{-7.1pt}\includegraphics[height=9pt]{head_foot/LF}}
\fancyfoot[CO]{\vspace{-7.1pt}\hspace{13.2cm}\includegraphics{head_foot/RF}}
\fancyfoot[CE]{\vspace{-7.2pt}\hspace{-14.2cm}\includegraphics{head_foot/RF}}
\fancyfoot[RO]{\footnotesize{\sffamily{1--\pageref{LastPage} ~\textbar  \hspace{2pt}\thepage}}}
\fancyfoot[LE]{\footnotesize{\sffamily{\thepage~\textbar\hspace{3.45cm} 1--\pageref{LastPage}}}}
\fancyhead{}
\renewcommand{\headrulewidth}{0pt} 
\renewcommand{\footrulewidth}{0pt}
\setlength{\arrayrulewidth}{1pt}
\setlength{\columnsep}{6.5mm}
\setlength\bibsep{1pt}
%%%END OF FOOTER%%%

%%%FIGURE SETUP - please do not change any commands within this section%%%
\makeatletter 
\newlength{\figrulesep} 
\setlength{\figrulesep}{0.5\textfloatsep} 

\newcommand{\topfigrule}{\vspace*{-1pt}% 
\noindent{\color{cream}\rule[-\figrulesep]{\columnwidth}{1.5pt}} }

\newcommand{\botfigrule}{\vspace*{-2pt}% 
\noindent{\color{cream}\rule[\figrulesep]{\columnwidth}{1.5pt}} }

\newcommand{\dblfigrule}{\vspace*{-1pt}% 
\noindent{\color{cream}\rule[-\figrulesep]{\textwidth}{1.5pt}} }

\makeatother
%%%END OF FIGURE SETUP%%%

%%%TITLE, AUTHORS AND ABSTRACT%%%
\twocolumn[
  \begin{@twocolumnfalse}
{\includegraphics[height=30pt]{head_foot/journal_name}\hfill%
 \raisebox{0pt}[0pt][0pt]{\includegraphics[height=55pt]{head_foot/RSC_LOGO_CMYK}}%
 \\[1ex]%
 \includegraphics[width=18.5cm]{head_foot/header_bar}}\par
\vspace{1cm}
\sffamily
\begin{tabular}{m{4.5cm} p{13.5cm} }

\includegraphics{head_foot/DOI} & \noindent\LARGE{\textbf{A high throughput search of efficient thermoelectric half-Heusler compounds$^\dag$}} \\%Article title goes here instead of the text "This is the title"
\vspace{0.3cm} & \vspace{0.3cm} \\

& \noindent\large{Parul R. Raghuvanshi\textit{$^{a}$}, Suman Mondal\textit{$^{a}$}, and Amrita Bhattacharya$^{\ast}$\textit{$^{a}$}} \\%Author names go here instead of "Full name", etc.

\end{tabular}

 \end{@twocolumnfalse} \vspace{0.6cm}

  ]
%%%END OF TITLE, AUTHORS AND ABSTRACT%%%

%%%FONT SETUP - please do not change any commands within this section
\renewcommand*\rmdefault{bch}\normalfont\upshape
\rmfamily
\section*{}
\vspace{-1cm}

%%%FOOTNOTES%%%

%\footnotetext{\textit{$^{a,b}$~Department of metallurgical engineering and materials science, Indian Institute of Technology, Bombay, Powai-400076, Mumbai, Maharashtra, India. E-mail: parul\_raghuvanshi@iitb.ac.in}}
%Please use \dag to cite the ESI in the main text of the article.
%If you article does not have ESI please remove the the \dag symbol from the title and the footnotetext below.
%additional addresses can be cited as above using the lower-case letters, c, d, e... If all authors are from the same address, no letter is required

%%%END OF FOOTNOTES%%%

%%%MAIN TEXT%%%%
\renewcommand{\thesection}{S\arabic{section}}
\renewcommand{\thefigure}{S\arabic{figure}}
\renewcommand{\thetable}{S\arabic{table}}

\section{Stability}
In order to examine the stability of the compositions, the formation energy of the compounds is calculated as;

\begin{equation}
\label{Eform}
E_\mathrm{f} = E(X_{m}Y_{n}Z_{t})-m \cdot \frac{E(X_{a})}{a}-n \cdot \frac{E(Y_{b})}{b}- t \cdot \frac{E(Z_{c})}{c}
\end{equation}
where, $E(X_{m}Y_{n}Z_{t})$ is the total energy of the compound and $E(X_{a})$, $E(Y_{b})$, and $E(Z_{c})$ are the total energy of the bulk phases of X, Y, and Z with $a$, $b$ and $c$ numbers of atoms respectively. 

The effect of temperature on formation energy is incorporated by adding the contributions from the vibrational free energy $F_\mathrm{vib}$ calculated using phonopy package:
\begin{equation}
\label{Ef}
E_\mathrm{f}(T) = E_\mathrm{f}  + F_\mathrm{vib}(T)
\end{equation}

The second term, $ F_\mathrm{vib}(T)$, is the free-energy contribution of the individual compositions appearing in eqn.~\ref{Eform},

\begin{equation}
\label{Evib}
F_\mathrm{vib}(T) = F(X_{m}Y_{n}Z_{t})-m \cdot \frac{F(X_{a})}{a}-n \cdot \frac{F(Y_{b})}{b}- t \cdot \frac{F(Z_{c})}{c}
\end{equation}

\begin{figure}
\centering
\includegraphics[scale=0.38]{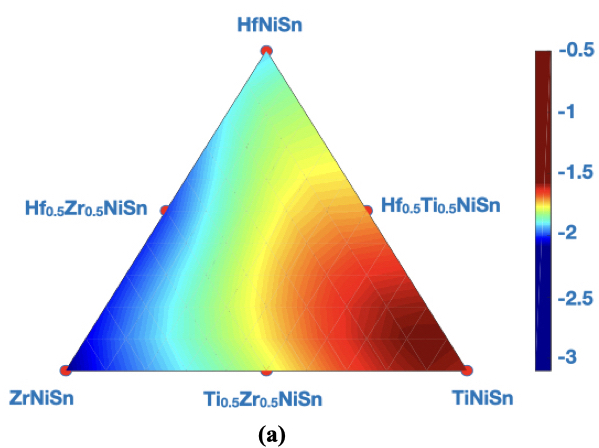}
\includegraphics[scale=0.38]{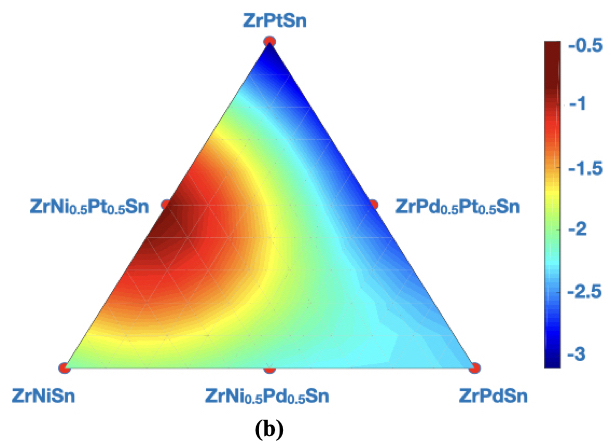}
\includegraphics[scale=0.38]{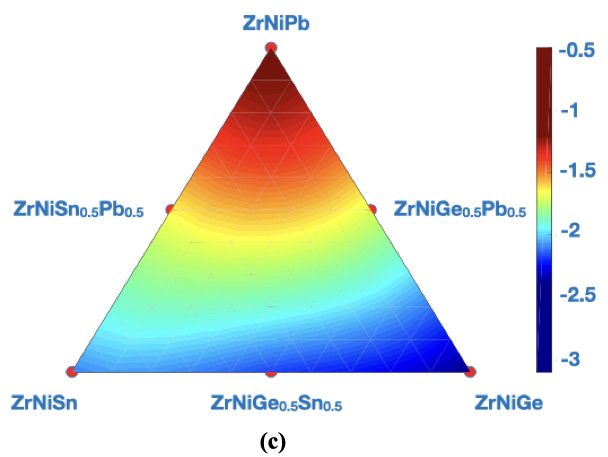}
\caption{Formation energy of X$_{p}$X$'_{1-p}$Y$_{q}$Y$'_{1-q}$Z$_{r}$Z$'_{1-r}$ for $p, q, r = 0, 0.5, 1$, calculated by incorporating the static energy as well as the vibrational free energy at 300 K for (a) X, X$'$ = Ti, Zr, Hf; Y= Ni; Z= Sn, (b) X= Zr; Y, Y$'$ = Ni, Pd, Pt; Z= Sn, and (c) X = Zr; Y= Ni; Z= Ge, Sn, Pb.}
\label{fig:FE}
\end{figure}

In order to explore the formability of the compounds, the formation energy~($E_{\mathrm{f}}$) of the compositions is calculated for all pristine compounds and their 50\% substituted counterparts using eqn.~\ref{Eform}. In order to explore the effect of temperature on the stability of the compounds, the contribution stemming from the vibrational free energy at 300 K of the compositions are incorporated in the formation energy equation~(cf. eqn.~\ref{Evib}). In all cases, the formation energy of the substituted counterparts are found to be comparable to the pristine ones and have negative formation energy, as shown in Fig.~\ref{fig:FE}, which indicates that all these structures can be realised in the laboratory.

\section{Degeneracy}
\begin{figure}
\centering
\centerline{\includegraphics[scale=.45]{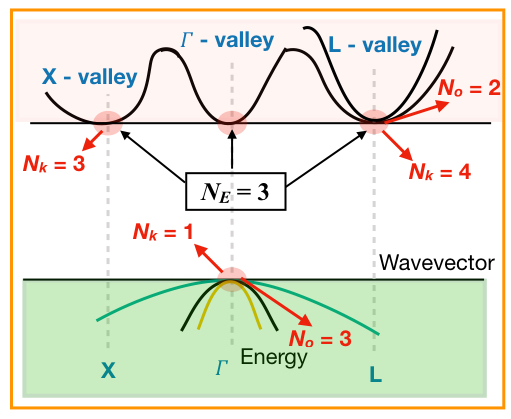}}
\caption{Schematic explaining the nomenclature for different degeneracies in an ideal case of face centered cubic system ($N_{k}$ : kpoint degeneracy, $N_{o}$ : orbital degeneracy and $N_{E}$ : valley degeneracy).}
\label{fig:deg}
\end{figure}

The transport coefficients $S$ and $\sigma$ depends on the microscopic parameter, viz. density of state effective mass, $m_{D}^{*}$ as,
\begin{equation}
S=\frac{8\pi ^{2}\kappa _{B}^{2}}{3eh^{2}}m_{D}^{*}T\left ( \frac{\pi}{3n} \right )^{2/3}
\label{equ:S}
\end{equation}

\begin{equation}
\sigma = \frac{n{e}^{2}\tau}{m^{*}} =ne\mu
\label{equ:1}
\end{equation}

where, $n$ is the carrier concentration, $\tau$ is the relaxation time and $\mu$ is the mobility. Further, $m_{D}^{*}$ can be written in terms of degeneracy ($N_{v}$) and the band effective mass ($m_{b}$) as,
 \begin{equation}
 m_{D}^{*} = N_{v}^{2/3} m_{b}^{*}
\label{equ:deg}
\end{equation}

where $N_{v}$ is the band degeneracy. To understand the role of $N_{v}$ in band structure schematic is shown in~Fig.\ref{fig:deg}. In which graphical representations of the degeneracies with our own set of nomenclature (keeping the existing literature in mind) has been provided. Fig.~\ref{fig:deg} represents the ideal case of fcc system with X-$\Gamma$-L wavevector path. Here, three valleys contributes in the CBm at three different high symmetry points (i.e., X, $\Gamma$, and L). Therefore, the figure shows the presence of a high number of degeneracies in the ideal structure and the value of $m_{D}^{*}$ of such structure can be expressed as:
 \begin{equation}
 m_{D}^{*} = N_{v1}^{2/3} m_{b1}^{*} + N_{v2}^{2/3} m_{b2}^{*} + N_{v3}^{2/3} m_{b3}^{*}
\label{equ:deg}
\end{equation}
where, $N_{v1}$, $N_{v2}$, and $N_{v3}$ are the degeneracies at three CBm valleys with their respective band effective masses. Hence, the value of density of state effective mass for the ideal fcc case comes out to be very high. 

\section{Parent half Heuslers}

\begin{table}
\begingroup
\caption{\label{tab:parent}Optimized lattice parameter (\AA) and band gap (eV) of parent half Heusler compounds.}
\renewcommand{\arraystretch}{1.1}
\begin{tabular}{cccccc}
\hline
System& Lattice parameter & Band gap & \\
\hline
%&Gap&0.49&0.40&0.20&0.54&0.45&\\
%\hline
HfNiSn         &  4.3199                    &  0.350    &\\
HfNiGe        & 4.1372                     & 0.561   & \\
HfPdGe        & 4.3285                     & 0.569   & \\
HfPdSn        & 4.4860                     & 0.393   & \\
HfPtGe        & 4.3560                     & 1.137    &\\
HfPtPb        & 4.5694                     & 0.751   & \\
HfPtSn        & 4.5011                     & 0.907   & \\
TiNiSn        &  4.2042                     &   0.400   &\\
TiNiGe        & 4.0036                     & 0.622   & \\
TiPdGe        & 4.2088                     & 0.662   & \\
TiPdSn        & 4.3852                     & 0.478    &\\
TiPtGe        & 4.2302                     & 0.866   & \\
TiPtSn        & 4.3974                     & 0.781   & \\
ZrNiGe       &4.1717                     &0.600  & \\
ZrNiSn       &4.3498                     &0.480 & \\
ZrNiPb       & 4.1717                    &0.620 & \\
ZrPdGe        & 4.3632                     & 0.638   & \\
ZrPdSn        &4.5165                      &0.430  & \\
ZrPdPb        & 4.3498                     & 0.424   & \\
ZrPtGe        & 4.3882                     & 1.189   & \\
ZrPtSn         &4.5314                     & 0.980   & \\
ZrPtPb        & 4.5985                     & 0.823  &\\
\hline
\end{tabular}
\endgroup
\end{table}

\begin{figure}[H]
\centering
\centerline{\includegraphics[scale=.35]{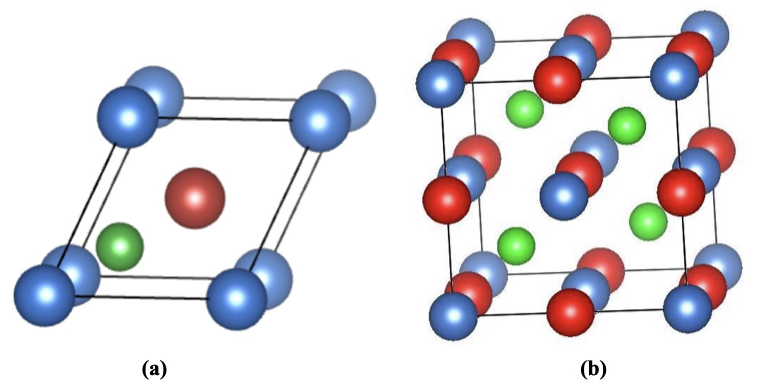}}
\caption{The crystal structure of (a) primitive cell, and (b) conventional cell of ZrNiSn. The blue, green and red represents Zr , Ni, and Sn atoms, respectively. }
\label{fig:conv}
\end{figure}

\begin{figure}
\centering
\centerline{\includegraphics[scale=.4]{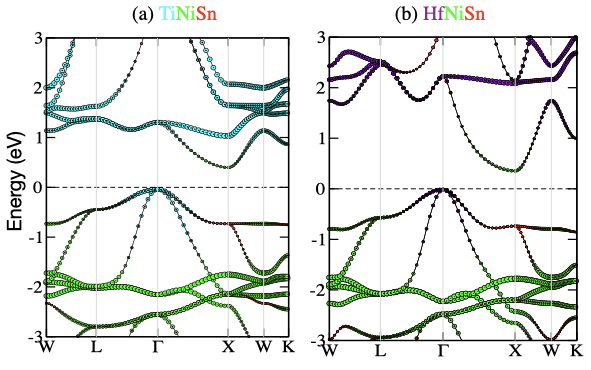}}
\centerline{\includegraphics[scale=.4]{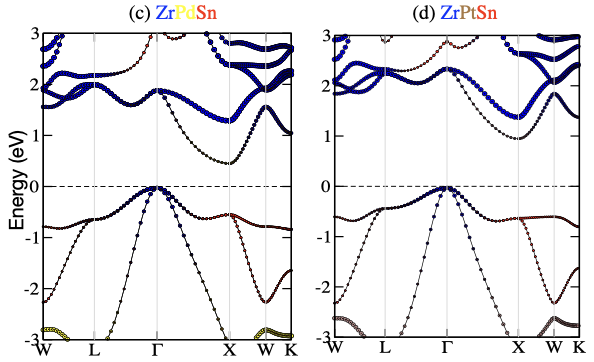}}
\centerline{\includegraphics[scale=.4]{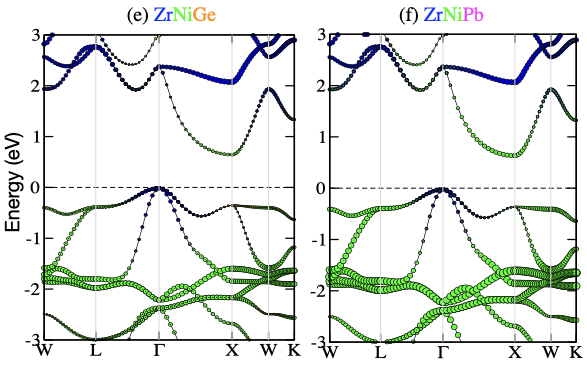}}
\caption{The electronic band structure (GGA-PBE) few of the half-Heusler compounds (a) TiNiSn, (b) HfNiSn, (c) ZrPdSn, (d) ZrPtSn, (e) ZrNiGe and (d) ZrNiPb.}
\label{fig:sband}
\end{figure}

Electronic band-structure calculations are carried out for 27 parent half-Heusler compounds.~Fig.\ref{fig:sband} shows the band-structure for 6 combination of such HH compounds with two different elements substituted each at X, Y and Z sites respectively for XYZ based compounds (where X= Ti, Hf; Y = Pd, Pt and Z = Ge, Pb). The colour codes used in these plots are same as that given in the primitive cell structure (cf.~Fig.2. in the manuscript). Calculations show that all the site substituted variant of ZrNiSn half-Heusler compounds have indirect band gap in the range 0.35 -1.12 eV. The relaxed lattice parameters and band gaps are shown in table S~\ref{tab:parent}.\\
The crystal structure of the conventional ZrNiSn containing 12 atoms (four Zr atoms, four Ni atoms and four Sn atoms at position (0, 0, 0), (1/4, 1/4, 1/4), and (1/2, 1/2, 1/2) respectively) in unit cell is shown in Fig.~\ref{fig:conv}

\subsection{Energy eigenvalues}
The tabulated values in table S~\ref{tab:eigenCBm}~and S\ref{tab:eigenVBM} are the energy eigenvalues (corresponding to the Fig. 4 (b), (c), (d), and (e) in the manuscript) at conduction band extrema and valence band extrema of respectively.
\begin{table}[H]
\centering
\begingroup
\caption{\label{tab:eigenCBm}The energy eigenvalues of CBm in units of eV}
\setlength{\tabcolsep}{3pt}
\begin{tabular}{ccccccccc}
\hline
System & CBm ($\Gamma$) & & CBm (M) & \hspace*{0.7mm} CBm+1 (M) \\
\hline
Ti$_{0.5}$Zr$_{0.5}$NiSn & 0.441 & & 0.439 & 0.439\\
ZrNi$_{0.5}$Pd$_{0.5}$Sn &  0.324 & & 0.174 & 0.522 \\
ZrNiGe$_{0.5}$Sn$_{0.5}$ &  0.509 & & 0.514 & 0.514 \\
Ti$_{0.5}$Zr$_{0.5}$NiGe$_{0.5}$Sn$_{0.5}$ & 0.429 & & 0.462 & 0.462 \\\hline
\end{tabular}
\endgroup
\end{table}

\begin{table}[H]
\centering
\begingroup
\caption{\label{tab:eigenVBM}The energy eigenvalues of VBM in units of eV}
\setlength{\tabcolsep}{3pt}
\begin{tabular}{ccccccccc}
\hline
System & VBM ($\Gamma$) & VBM-1 ($\Gamma$)& VBM-2 ($\Gamma$) \\
\hline
Ti$_{0.5}$Zr$_{0.5}$NiSn & -0.002 &  -0.002 &  -0.002 \\
ZrNi$_{0.5}$Pd$_{0.5}$Sn & -0.001 & -0.117& -0.117 \\
ZrNiGe$_{0.5}$Sn$_{0.5}$ &  -0.001 & -0.001& -0.001 \\
Ti$_{0.5}$Zr$_{0.5}$NiGe$_{0.5}$Sn$_{0.5}$ & -0.004 & -0.047& -0.047 \\\hline
\end{tabular}
\endgroup
\end{table}

\subsection{Spin-orbit coupling (soc)}
\begin{figure}[H]
\centering
\centerline{\includegraphics[scale=.41]{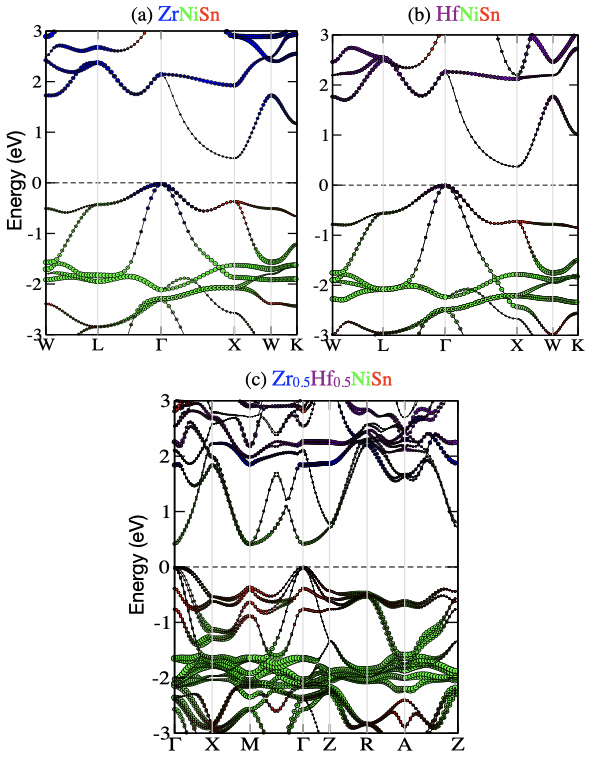}}
\caption{The electronic band structure (GGA-PBE including the spin orbit correction) of the respective cases of (a) ZrNiSn, (b) HfNiSn, and (c) Zr$_{0.5}$Hf$_{0.5}$NiSn. Zr (blue), Hf (purple), Ni (green), and Sn (red).}
\label{fig:socband}
\end{figure}

The bandstructures shown in Fig.~\ref{fig:socband} (a), (b), and (c) are after the inclusion of spin orbit coupling (soc). It is clear from the figure that the degeneracies and band pattern are same as that of non-soc HH compounds. However, due to soc inclusion a small effect is visible on the band gap of ZrNiSn ($\sim$ 0.48 eV), while a distinct effect on the band gap of HfNiSn ($\sim$ 0.32 eV), as shown in Fig.~\ref{fig:socband} (a) and (b) respectively. Similarly, band gap reduces a bit for the 50\% substituted compound Zr$_{0.5}$Hf$_{0.5}$NiSn (E$_{g}$ $\sim$ 0.40 eV) as shown in Fig.~\ref{fig:socband} (c). 

\subsection{Structure of ZrNiSn}
\begin{figure}[H]
\centering
\centerline{\includegraphics[scale=.42]{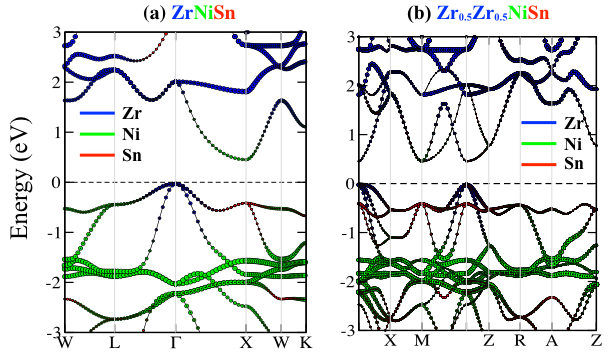}}
\caption{The electronic band structure (GGA-PBE) of (a) ZrNiSn, and (b) Zr$_{0.5}$Zr$_{0.5}$NiSn. Zr (blue),  Ni (green), and Sn (red).}
\label{fig:compare}
\end{figure}

The electronic band structure is compared for primitive and in the conventional tetragonal lattice cell of ZrNiSn is shown in Fig~\ref{fig:compare}. Please note that the electronic band structure of the parent XYZ (in the tetragonal high symmetry path) compound, is similar to the (X and Z site) substituted ones reported for 50 \% doping in the manuscript in all regards in terms of valley and tip degeneracies. However, the difference in transport coefficients arises from the difference in the curvature of the bands etc. 

\subsection{X site substitution}
\label{sec:sX}
\begin{figure}[H]
\centering
\centerline{\includegraphics[scale=.42]{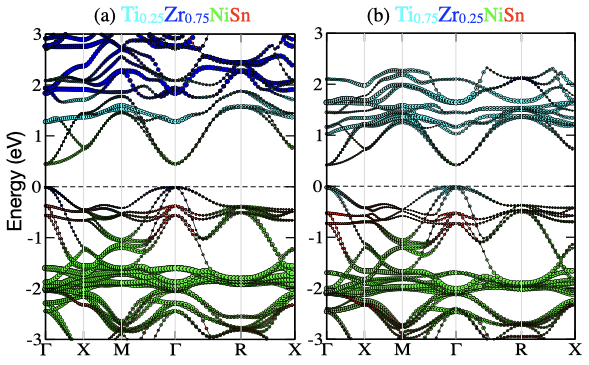}}
\caption{The electronic band structure (GGA-PBE) of 0.25 and 0.75 composition substitution at X site (a) Ti$_{0.25}$Zr$_{0.75}$NiSn, and (b) Ti$_{0.75}$Zr$_{0.25}$NiSn. Zr (blue), Ti (cyan),  Ni (green), and Sn (red).}
\label{fig:xband}
\end{figure}

The Zr (or Ti) and Ni d orbital interaction leads to a formation of band gap due to contribution of VBM and CBm as shown by blue (cyan) and green colours respectively in~Fig.\ref{fig:xband}. Specifically, conduction band edge is formed due to the Ni-d character and valence band edge is formed as a result of Zr-d character which is clearly depicted in all the figures. Whereas, Sn (s-p orbitals) contribute in the lower energy valence bands (red coloured bands). Furthermore, to clearly understand the states near the Fermi level, we analyzed the bandstructure for all the X-site substituted cases (X$_{p}$X$'_{1-p}$NiSn, where, X, X$'$=Ti-Zr, Zr-Hf and Hf-Ti, $p$ = 0.25, 0.5, 0.75) and it is observed that lighter element contribute more at the valence band edge as shown in Fig.~\ref{fig:xband}.

The substitution of 25\% and 75\% Ti in ZrNiSn fold the bands when going from face centered primitive to conventional cell, and become direct band gap semiconductor with all the maxima and minima positioned at $\Gamma$ point as shown in Fig.~\ref{fig:xband}. The bandstructure shows not much difference in band gap for both the cases of substitution, where E$_{g}$ is 0.42 eV  and 0.40 eV for Ti$_{0.25}$Zr$_{0.75}$NiSn and Ti$_{0.75}$Zr$_{0.25}$NiSn respectively.\\

\subsection{Y site substitution}
\label{sec:sY}
The electronic band structure and the partial density of states (PDOS) of ZrNiSn are calculated, and found that the orbital contribution to the conduction band (CB) is formed due to the interaction between Zr and Ni orbitals. 

\begin{figure}[H]
\centering
\centerline{\includegraphics[scale=.3]{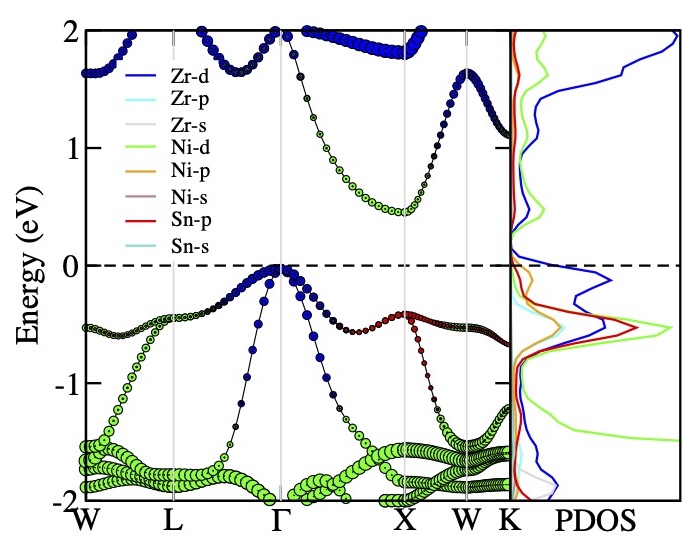}}
\caption{The electronic band structure and projected density of states of primitive ZrNiSn. Different colors representing s, p, and d orbital contribution of Zr, Ni, and Sn.}
\label{fig:Y1}
\end{figure}

\begin{figure}[H]
\centering
\centerline{\includegraphics[scale=.35]{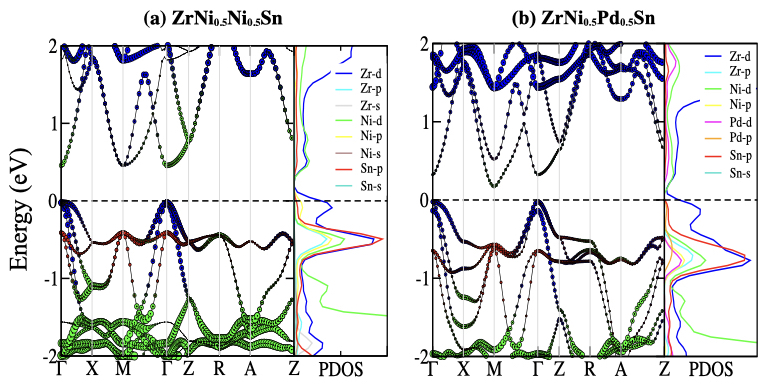}}
\caption{The electronic band structure and projected density of states of ZrNi$_{0.5}$Ni$_{0.5}$Sn and ZrNi$_{0.5}$Pd$_{0.5}$Sn. Different colors representing s, p, and d orbital contribution of Zr, Ni (Pd), and Sn.}
\label{fig:Y2}
\end{figure}

\begin{figure}[H]
\centering
\centerline{\includegraphics[scale=.3]{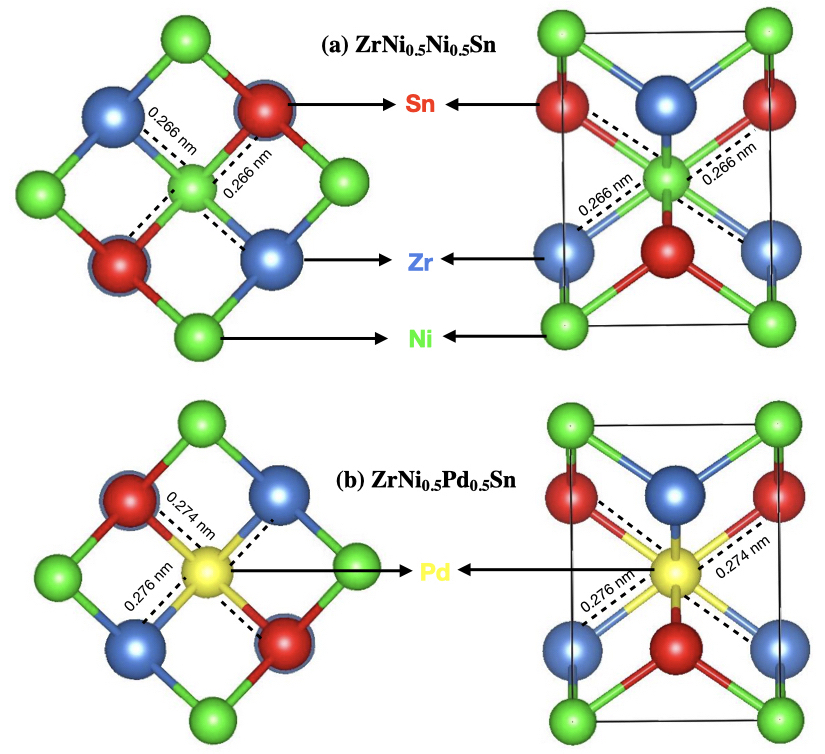}}
\caption{The crystal structure in planar view and side view of (a) ZrNi$_{0.5}$Ni$_{0.5}$Sn, and (b) ZrNi$_{0.5}$Pd$_{0.5}$Sn. The blue, green (yellow) and red represents the Zr, Ni (Pd), and Sn atoms, respectively. Bond lengths (in nm) are shown as dashed lines.}
\label{fig:Y3}
\end{figure}

Thus, it is apparent that substitution at X site can adjust the band structure near the band edge and at Y site substitution will affect the bottom of the CB. 
In order to investigate the reason behind the breaking of the degenerate bands at the M point upon 50 \% substitution at the Y site, the eBS and pDOS of ZrNi$_{0.5}$Pd$_{0.5}$Sn has been investigated and compared with ZrNi$_{0.5}$Ni$_{0.5}$Sn by relaxing the structure in the tetragonal geometry as shown in the Fig.~\ref{fig:Y2}. From the partial DOS the bottom of the CB was found to have mixed contribution coming from all the d-orbitals. Upon close investigation of the structure, a breaking in the local symmetry is observed for the Pd atom in the center, whereby the neighboring bond distance of Pd-Sn (0.274 nm) is found to be different from the Pd-Zr (0.276 nm). However, in case of Ni atom, the Ni-Zr and Ni-Sn bond distances are found to be the same (0.266 nm)  as shown in the Figure shown below Fig.~\ref{fig:Y3}. Similar situation was also found for the Pt substitution. 

\begin{figure}[H]
\centering
\centerline{\includegraphics[scale=.3]{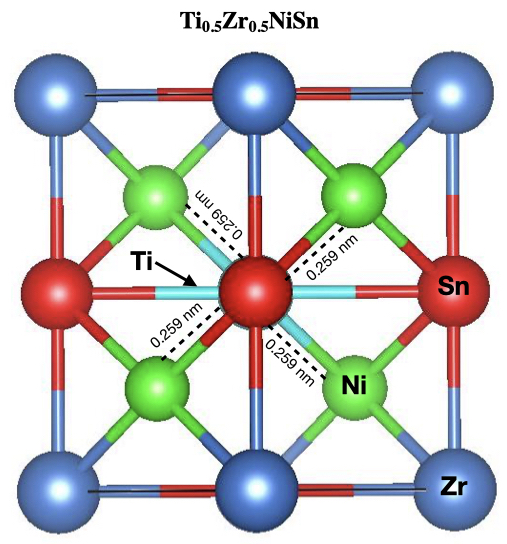}}
\caption{The crystal structure planar view of Ti$_{0.5}$Zr$_{0.5}$NiSn. The blue (cyan), green, and red represents the Zr (Ti), Ni, and Sn atoms, respectively. Bond lengths (in nm) are shown as dashed lines.}
\label{fig:Y4}
\end{figure}

However, such situation does not arise in X site as well as the Z site substitution with the substituted atom sitting at the center, in which the same elements are found to occupy the equivalent symmetric neighbor positions as shown in figure below Fig.~\ref{fig:Y4}. 

\begin{figure}[H]
\centering
\centerline{\includegraphics[scale=.42]{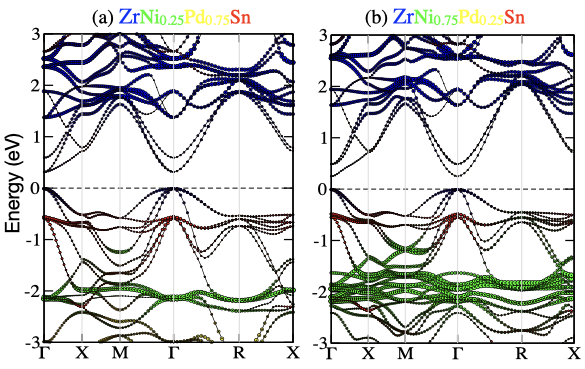}}
\caption{The electronic band structure (GGA-PBE) of 0.25 and 0.75 composition substitution at Y site (a) ZrNi$_{0.25}$Pd$_{0.75}$Sn, and (b) ZrNi$_{0.75}$Pd$_{0.25}$Sn. Zr (blue),  Ni (green), Pd (yellow) and Sn (red).}
\label{fig:yband}
\end{figure}

Similar to the 50\% substituted ZrNi$_{0.5}$Pd$_{0.5}$Sn compound (see Y-site substitution section), the splitting of bands at CBm is also notifying in the cases of 25\% and 75\%, Pd substitution at Ni site as shown in ~Fig.\ref{fig:yband} (a) and (b) respectively. The band gap is also narrow in both the ZrNi$_{0.25}$Pd$_{0.75}$Sn (E$_{g}$$\sim$ 3 eV) and ZrNi$_{0.75}$Pd$_{0.25}$Sn (E$_{g}$$\sim$ 0.25 eV) compounds, means biploar effect can also be dominate and diminish the transport properties in these compositons.

\subsection{Z site substitution}
\label{sec:sZ}
\begin{figure}[H]
\centering
\centerline{\includegraphics[scale=.42]{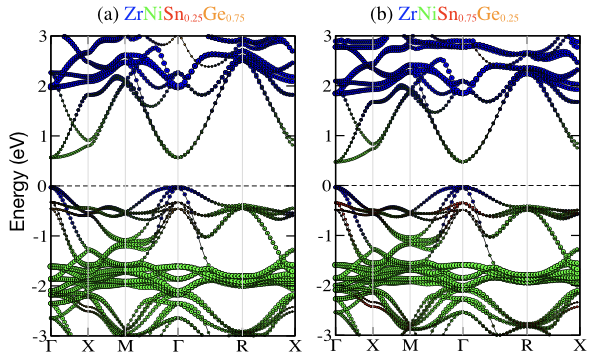}}
\caption{The electronic band structure (GGA-PBE) of 0.25 and 0.75 composition substitution at Z site (a) ZrNiGe$_{0.75}$Sn$_{0.25}$, and (b) ZrNiGe$_{0.25}$Sn$_{0.75}$. Zr (blue), Ni (green), Sn (red), and Ge (orange).}
\label{fig:zband}
\end{figure}

The 25\% and 75\% substitution at Z site is similar to the X site substituted compounds in terms of degeneracy. In both the cases maxima and minima lies at $\Gamma$ point with same number of total degeneracy. However, the band gap is affected more with the \% of Ge substitution at Sn site, and it increases to 0.56 eV  for ZrNiGe$_{0.75}$Sn$_{0.25}$ (cf.~Fig.~\ref{fig:zband} (a)) from 0.45 eV for ZrNiGe$_{0.25}$Sn$_{0.75}$ (cf.~Fig.~\ref{fig:zband} (b)). 

\section{Debye Callaway Formalism}
The modified Debye Callaway theory is used to model the scattering lifetimes and calculate the lattice thermal conductivity of the compounds. The scattering lifetimes namely the normal and Umklapp phonon scattering lifetime ${\tau_\mathrm{N}^{\mathrm{i}}}$ and ${\tau_\mathrm{u}^{\mathrm{i}}}$ respectively, corresponding to the longitudinal acoustic ($i$ = LA) and transverse modes ($i$ = TA). All the scattering lifetimes are written as a function of $x=\frac{\hbar \omega}{\mathrm{K_B}T} $

\begin{equation}
\begin{array}{lcl}
\label{tau_TA}
\displaystyle \frac{1}{\tau_\mathrm{N}^{\mathrm{TA}}(x)} =\frac{{\mathrm{K_B}}^4\gamma_{\mathrm{TA}}^2\mathrm{V}}{\mathrm{M}\mathrm{\hbar}^3 v_{\mathrm{TA}}^5}(\frac{\mathrm{K_B}}{\mathrm{\hbar}}) x T^5 
\end{array}
\end{equation}

\begin{equation}
\begin{array}{lcl}
\label{tau_LA}
\displaystyle \frac{1}{\tau_\mathrm{N}^{\mathrm{LA}}(x)} =\frac{{\mathrm{K_B}}^3\gamma_{\mathrm{LA}}^2 \mathrm{V}}{\mathrm{M}\mathrm{\hbar}^2 v_{\mathrm{LA}}^5}(\frac{\mathrm{K_B}}{\mathrm{\hbar}})^2 x^2 T^5 
\end{array}
\end{equation}

\begin{equation}
\begin{array}{lcl}
\label{tau_U}
\displaystyle \frac{1}{\tau_{\mathrm{U}}^{i}(x)} =\frac{\mathrm{\hbar}\gamma^2}{\mathrm{M} v_i^2 \theta_i}{(\frac{\mathrm{K_B}}{\mathrm{\hbar}})}^2 x^2 T^3 e^{-\theta_i/3T}.
\end{array}
\end{equation}

where, M is the average mass of an atom in the crystal, V is volume of the lattice, $\gamma$ is the average of the Gr$\ddot{\mathrm{u}}$neisen parameter ($\gamma=\frac{\sum \gamma_{i}} {\sum n_{i}}$) and $\theta_{i}$ is the Debye temperature, corresponding to the acoustic phonon mode $i$ ($i$ = LA, TA, TA$’$)

The lattice thermal conductivity $\kappa_{l}^{i}$ of the compounds is calculated from all the three 
scattering lifetimes, ${\tau_\mathrm{N}^{\mathrm{LA}}}$, ${\tau_\mathrm{N}^{\mathrm{TA/TA'}}}$ and ${\tau_\mathrm{u}^{\mathrm{i}}}$ of the acoustic phonon modes ($i$ = LA, TA, TA$’$)  using the equations (\ref{tau_TA} -\ref{tau_U}) as, 

\begin{equation}
\begin{array}{lcl}
\label{K}
\displaystyle {{\kappa_\mathrm{l}}^i} =\frac{1}{3}\mathrm{C}_iT^3\lbrace \int_0^{\theta_i/T} \frac{\tau_c^i x^4 e^x}{{(e^x -1)}^2}dx + \frac{[\int_0^{\theta_i/T} \frac{\tau_c^i x^4 e^x}{\tau_\mathrm{N}^i{(e^x -1)}^2}]^2}{\int_0^{\theta_i/T} \frac{\tau_c^i x^4 e^x}{\tau_\mathrm{U}^i\tau_\mathrm{N}^i{(e^x -1)}^2}}\rbrace \nonumber \\
\end{array}
\end{equation}

Where, $\mathrm{C}_i = \frac{{\mathrm{K_B}}^4}{2\pi^2 \mathrm{\hbar}^3 v_i}$ depends on the group velocity $v_i$ of the concerned mode and $\tau_c^i = \frac{\tau_\mathrm{U}^i \tau_\mathrm{N}^i} {\tau_\mathrm{U}^i+ \tau_\mathrm{N}^i}$ is the relaxation time. The total thermal conductivity is written as $\kappa_\mathrm{l}= {\kappa_{\mathrm{l}}^{\mathrm{TA}}} + {\kappa_{\mathrm{l}}^{\mathrm{LA}}}$.\\

\section{Transport coefficients}

\subsection{Electronic transport properties}
\begin{figure}[H]
\centering
\centerline{\includegraphics[scale=.2]{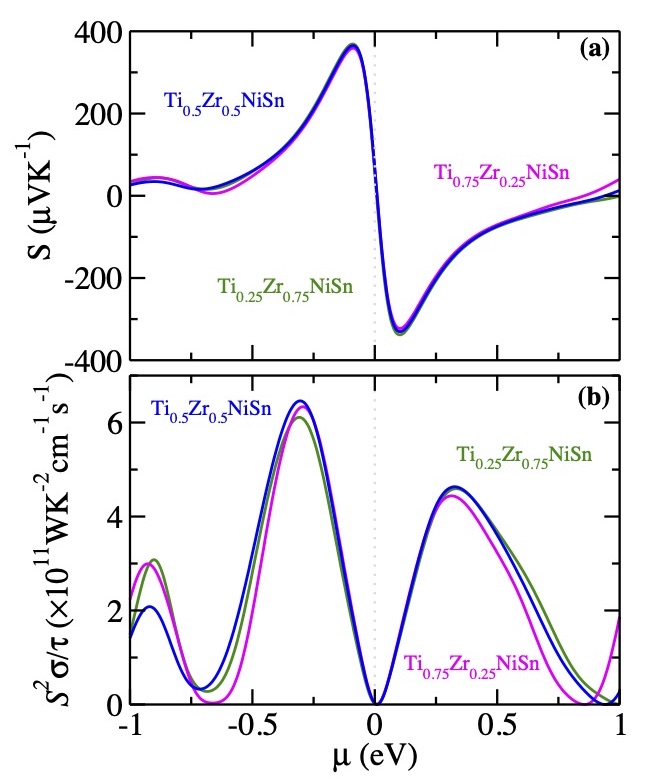}\hspace*{2mm}\includegraphics[scale=.2]{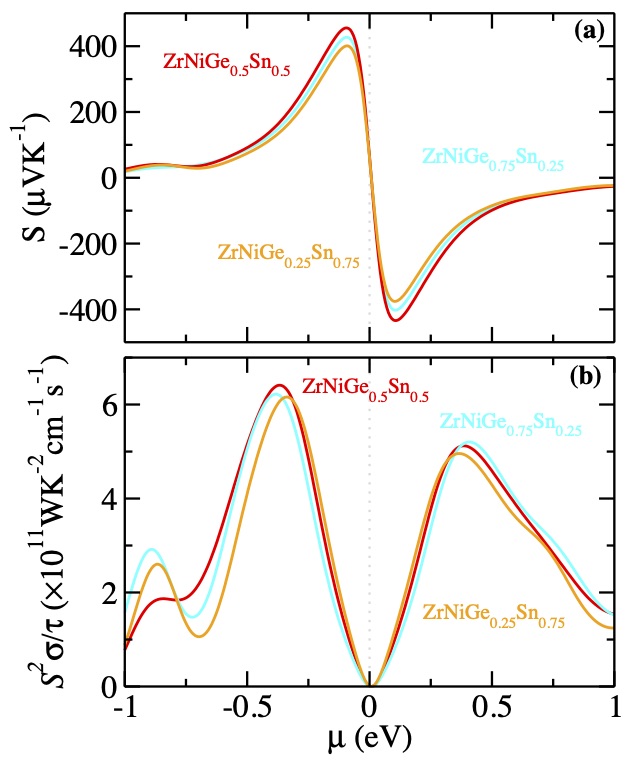}}
\caption{(a) Seebeck coefficient ($S$) and (b) power factor ($S^{2}\sigma/\tau$) of 0.25 and 0.75 composition substitution at X site (Ti$_{0.25}$Zr$_{0.75}$NiSn (green), and Ti$_{0.75}$Zr$_{0.25}$NiSn (magenta) compared with Ti$_{0.5}$Zr$_{0.5}$NiSn (blue)) and Z site (ZrNiGe$_{0.75}$Sn$_{0.25}$ (cyan), and ZrNiGe$_{0.25}$Sn$_{0.75}$ (orange) compared with ZrNiGe$_{0.5}$Sn$_{0.5}$ (red)) plotted as a function of electronic chemical potential $\mu$ for few selected substituted compounds.}
\label{fig:trans}
\end{figure}

The Seebeck coefficient (S) and power factor (PF) of 25 \% and 75 \% X and Z site substituted compounds found comparable to the 50 \% substituted compounds for both $n$- and $p$- type, as explicitly shown in Fig.~\ref{fig:trans}. However, an effective decrease in the lattice thermal conductivity may be obtained for 50\% substitution at any site as compared to the 25 \% and 75 \% substituted compositions empirically, due to the mass fluctuation scattering introduced by the mass difference at substituted sites using,
\begin{equation}
\label{eq:pd}
\tau_{D}^{-1}\propto A \omega ^{4 }
\end{equation}

Where, $A \omega ^{4 }$ represents the scattering by point impurities or isotopes. Where, A depends upon the mass and size contrast of the alloyed elements, $A\propto f_{1}f_{2}\left \{\left ( \frac{M_{1} - M_{2}}{M} \right )^2 + \left ( \frac{r_{1} - r_{2}}{r} \right )^2\right \}$, $M_{i}$ the mass of atoms, $r_{i}$ is radius and $f_{i}$ is fraction of atoms with mass $M_{i}$. For example, in Ti$_{0.75}$Zr$_{0.25}$NiSn, $f_{1}$= 0.25 and $f_{2}$ = 0.75, $A\propto 0.1875\left \{\left ( \frac{M_{1} - M_{2}}{M} \right )^2 + \left ( \frac{r_{1} - r_{2}}{r} \right )^2\right \}$, and Ti$_{0.5}$Zr$_{0.5}$NiSn, $f_{1}$= 0.5 and $f_{2}$ = 0.5, $A\propto 0.25\left \{\left ( \frac{M_{1} - M_{2}}{M} \right )^2 + \left ( \frac{r_{1} - r_{2}}{r} \right )^2\right \}$. Therefore, 0.25 and 0.75 compositional fractions lead to compromise with the lattice thermal conductivity, even after obtaining comparable PF. This empirical relation is also in agreement with our DFT results as Ti$_{0.25}$Zr$_{0.75}$NiSn ($\kappa_{l}$ = 3.6) and ZrNiGe$_{0.25}$Sn$_{0.75}$ (($\kappa_{l}$ = 3.2)) showed higher lattice thermal conductivity as compared to their respective 50\% substituted compounds i.e. Ti$_{0.5}$Zr$_{0.5}$NiSn ($\kappa_{l}$ = 2.8) and ZrNiGe$_{0.5}$Sn$_{0.5}$ ($\kappa_{l}$ = 3.1).

\subsubsection{Anisotropy in the electronic transport coefficients}
\begin{figure}[H]
\centering
\centerline{\includegraphics[scale=.28]{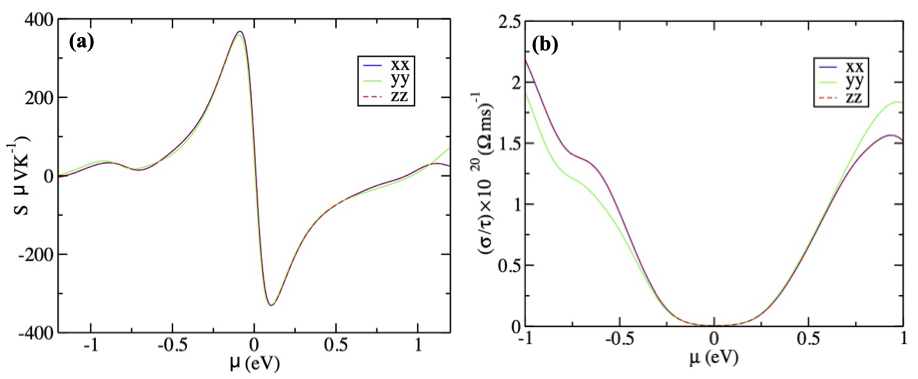}}
\caption{Diagonal terms of the (a) Seebeck and (b) electrical conductivity tensors vs the chemical potential at 700 K for the tetragonal Ti$_{0.5}$Zr$_{0.5}$NiSn compound.}
\label{fig:diagonal}
\end{figure}

We have investigated the anisotropy in $\sigma/\tau$ and S for the compounds. As expected, for the parent cubic compounds, the $\sigma/\tau$ and S are found to be isotropic with equal diagonal elements in their conductivity and Seebeck tensor i.e. with equal xx, yy and zz components. 
However, the substituted compounds with tetragonal lattice i.e. X$_{0.5}$X'$_{0.5}$NiSn showed anisotropy i.e. $\frac{\sigma _{xx}}{\tau}= \frac{\sigma _{zz}}{\tau}\neq \frac{\sigma _{yy}}{\tau}$ for the $\sigma$ and similarly for S (see Fig.~\ref{fig:diagonal} below for Ti$_{0.5}$Zr$_{0.5}$NiSn compound). This is in agreement with the results of the effective mass in Table 2 of main manuscript which is found to be anisotropic along different directions. However, no strong anisotropy is observed in any of the reported compound along different directions as can be seen in Fig.~\ref{fig:diagonal}.

\subsection{Band dispersion}
\begin{figure}[H]
\centering
\centerline{\includegraphics[scale=.37]{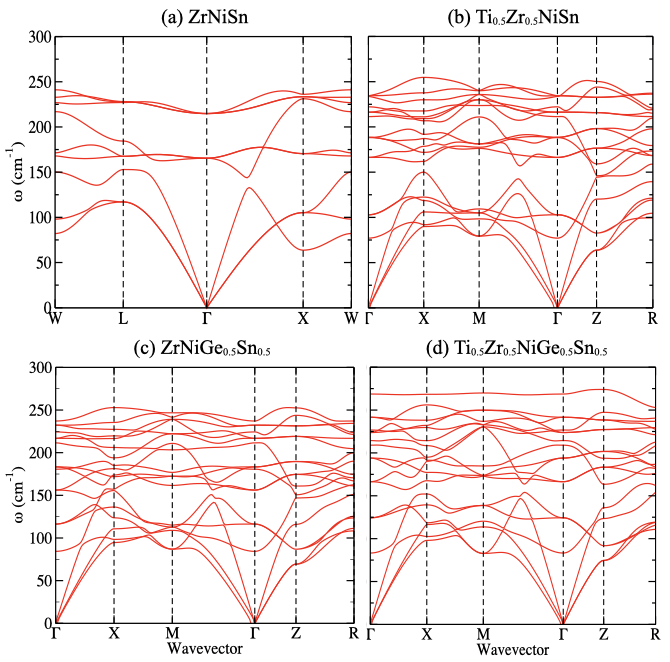}}
\caption{Phonon dispersion of a respective case of X, Z and co-site doping (a) ZrNiSn, (b) Ti$_{0.5}$Zr$_{0.5}$NiSn, (c) ZrNiGe$_{0.5}$Sn$_{0.5}$, and (d) Ti$_{0.5}$Zr$_{0.5}$NiGe$_{0.5}$Sn$_{0.5}$ along the Wavevector.}
\label{fig:trans1}
\end{figure}

The harmonic phonon dispersion of parent, X-site, Z-site and co-site substituted compound ZrNiSn, Ti$_{0.5}$Zr$_{0.5}$NiSn, ZrNiGe$_{0.5}$Sn$_{0.5}$ and Ti$_{0.5}$Zr$_{0.5}$NiGe$_{0.5}$Sn$_{0.5}$ respectively along the high symmetry path in the Brillouin zone is shown in Fig. \ref{fig:trans1}.

\subsection{Gr$\ddot{\mathrm{u}}$neisen parameter}
\begin{figure} [H]
\centering
\centerline{\includegraphics[scale=.3]{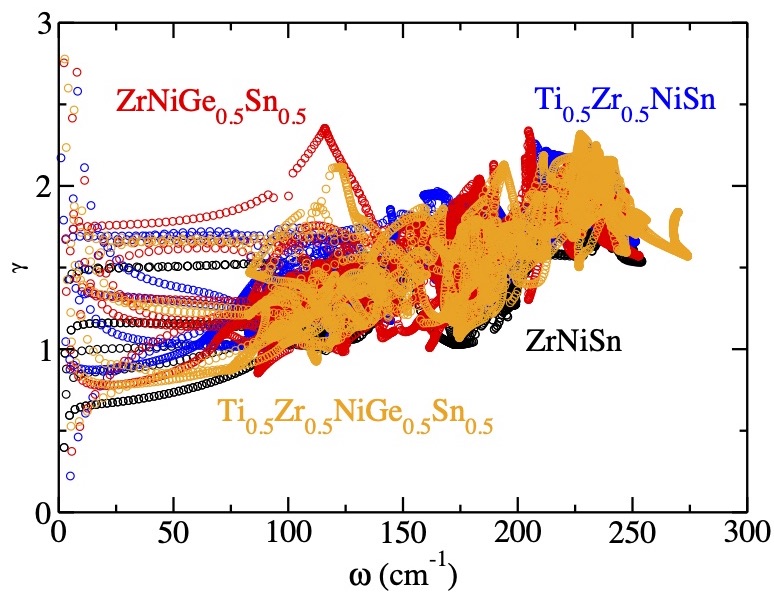}}
\caption{The comparison of Gr$\ddot{\mathrm{u}}$neisen  parameter ($\gamma$) of ZrNiSn, Ti$_{0.5}$Zr$_{0.5}$NiSn, ZrNiGe$_{0.5}$Sn$_{0.5}$, and Ti$_{0.5}$Zr$_{0.5}$NiGe$_{0.5}$Sn$_{0.5}$ as a function of frequency.}
\label{fig:trans2}
\end{figure}

The dispersion of the Gr$\ddot{\mathrm{u}}$neisen parameter along the high symmetry path is plotted in the Fig.~\ref{fig:trans2}. The phonon scattering lifetimes $\tau$s are inversely proportional to the square of the Gr$\ddot{\mathrm{u}}$neisen parameter~$\gamma^2$~(see.~Eqn.~\ref{tau_TA} -\ref{tau_U}). It is clearly shown in Fig.~\ref{fig:trans2} that $\gamma$ is higher for the substituted compounds as compared to the pristine ZrNiSn.  Therefore, Z-site substituted compound ZrNiGe$_{0.5}$Sn$_{0.5}$ has lowest $\kappa_\mathrm{l}$ in single site substituted and Ti$_{0.5}$Zr$_{0.5}$NiGe$_{0.5}$Sn$_{0.5}$ in co-site substituted compounds (high $\gamma$ signifies low $\tau$ and hence low $\kappa_\mathrm{l}$).

\subsection{Group velocity}
\begin{figure}[H]
\centering
\centerline{\includegraphics[scale=.3]{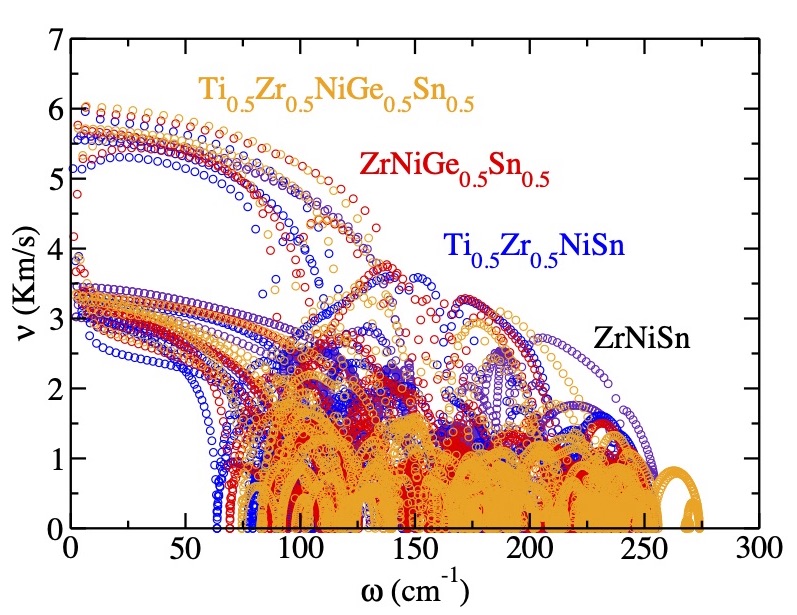}}
\caption{The comparison of group velocity ($v$) of ZrNiSn, Ti$_{0.5}$Zr$_{0.5}$NiSn, ZrNiGe$_{0.5}$Sn$_{0.5}$, and Ti$_{0.5}$Zr$_{0.5}$NiGe$_{0.5}$Sn$_{0.5}$ as a function of frequency.}
\label{fig:trans3}
\end{figure}

Similar to Gr$\ddot{\mathrm{u}}$neisen parameter $\tau$s are effected by the group velocities $v_i$s.  $\tau$s are directly proportional to the powers of $v_i$s~(cf.~Eqn.~\ref{tau_TA} -\ref{tau_U}), i.e., high $v_i$ indicates high $\tau$ and hence, high $\kappa_\mathrm{l}$. The group velocity $v$ is plotted as a function of frequency in Fig.~\ref{fig:trans3}, which is comparable for both the single site substituted compounds and also co-site compound. $v_\mathrm{TA/LA}$ is calculated from the maximum group velocity attained by the phonons at the $\Gamma$ point.

\subsection{Point defect scattering}
According to the Matthiessen’s rule the combined relaxation time $\tau_{C}$ can be expressed by the addition of the inverse relaxation times for the different scattering processes as, $\tau_{C}^{-1}=\sum _{i}\tau _{i}^{-1}$  , where the parameters $\tau_{i}$ are for the phonon relaxation time for the different scattering processes. Whereas, the lattice thermal conductivity has contributions stemming from different scattering processes such as, the point impurity/defect scattering, acoustic and Umklapp scattering (phonon phonon) and boundary scattering etc. (see reference 60 of the manuscript) and can be written as,

\begin{equation}
\begin{array}{lcl}
\label{masstau}
\tau_{u}^{-1}=A\omega ^{4}+B_{1}T^{3}\omega ^{2}+c/L
\end{array}
\end{equation}

The Debye Callaway model for the lattice thermal conductivity used in our study, however, is only limited to the acoustic and Umklapp phonon scattering processes. The point defect scattering is not explicitly modelled in this study and may have positive influence in the lattice thermal conductivity if taken into account.    

In the point defect scattering, the strain and mass differences between the substitutional and host atoms act as scattering centres for phonons and represents as point defects disorder. Thus, the point defects arise from both the mass and strain differences within the lattice, which is given as, 

\begin{equation}
\begin{array}{lcl}
\label{masstau2}
\tau_{D}^{-1}=\tau_{M}^{-1}+\tau_{S}^{-1}=\frac{V\omega ^{4}}{4\pi\nu_{s} ^{3}}(\Gamma_{M}+\Gamma_{S})
\end{array}
\end{equation}

Where, $\Gamma_{M}$  and $ \Gamma_{S}$   are the disorder scattering parameters due to mass and strain fluctuations respectively and $\nu_{s}$ is the average phonon group velocity. Moreover, for the simplicity eqn.~\ref{masstau2} can also be written as eqn.~\ref{eq:pd}.

As for instance, let us limit to the case of the disordered compounds viz. Ti$_{0.5}$Zr$_{0.5}$NiSn and Ti$_{0.25}$Zr$_{0.75}$NiSn. In this case, the disorder arises due to the presence of both Ti and Zr atoms in the Zr sublattice (i.e. in its ordered counterpart ZrNiSn). As the radius difference between Ti and Zr is very small, it may be assumed that the majority of phonon scattering is arising from the mass fluctuation scattering and not the strain. Hence, $\Gamma_{Total}\approx \Gamma_{M}\approx \frac{1}{3}\left (\frac{\bar{M}_{1}}{\bar{\bar{M}}}  \right )^{2}\Gamma_{M}^{i}$ (see reference 72 of the manuscript), which gives $\Gamma_{M} \approx 0.0228$ and $\Gamma_{M} \approx 0.0156$ for Ti$_{0.5}$Zr$_{0.5}$NiSn and Ti$_{0.25}$Zr$_{0.75}$NiSn compounds respectively. Thus, the disorder scattering parameter due to mass fluctuation is lower for 25\% substituted compound, we can justify that 50\% substituted compound has high mass fluctuation scattering effects of phonons than any other \% of substitution. 
This point defect phonon scattering is very dominant mechanism to describe the phonon scattering process in the solid solutions, which give rise to the reduction of lattice thermal conductivity $\kappa_{l}$. 

%%%END OF MAIN TEXT%%%


\balance


%%%REFERENCES%%%
\bibliography{rsc} %You need to replace "rsc" on this line with the name of your .bib file
\bibliographystyle{rsc} %the RSC's .bst file